\documentclass[preprint2]{proto}
\usepackage{times}
\usepackage{color}

\newcommand{\msun}{M$_\odot$}
\def\citeapos#1{\citeauthor{#1}'s (\citeyear{#1})} 
\begin{document}

\title{\textbf{\LARGE Angular momentum evolution of young low-mass stars and brown dwarfs: observations and theory}}

\author {\textbf{\large J\'er\^ome Bouvier}}
\affil{\small\em Observatoire de Grenoble}
\author {\textbf{\large Sean P. Matt}}
\affil{\small\em Exeter University}
\author {\textbf{\large Subhanjoy Mohanty}}
\affil{\small\em Imperial College London}
\author {\textbf{\large Aleks Scholz}}
\affil{\small\em University of St Andrews}
\author {\textbf{\large Keivan G. Stassun}}
\affil{\small\em Vanderbilt University}
\author {\textbf{\large Claudio Zanni}}
\affil{\small\em Osservatorio Astrofisico di Torino}

\begin{abstract}
\baselineskip = 11pt
\leftskip = 0.65in 
\rightskip = 0.65in
\parindent=1pc {\small This chapter aims at providing the most
  complete review of both the emerging concepts and the latest
  observational results regarding the angular momentum evolution of
  young low-mass stars and brown dwarfs. In the time since Protostars
  \& Planets V, there have been major developments in the availability
  of rotation period measurements at multiple ages and in different
  star-forming environments that are essential for testing theory. In
  parallel, substantial theoretical developments have been carried out
  in the last few years, including the physics of the star-disk
  interaction, numerical simulations of stellar winds, and the
  investigation of angular momentum transport processes in stellar
  interiors. This chapter reviews both the recent observational and
  theoretical advances that prompted the development of renewed
  angular momentum evolution models for cool stars and brown
  dwarfs. While the main observational trends of the rotational
  history of low mass objects seem to be accounted for by these new
  models, a number of critical open issues remain that are outlined in
  this review.  
  \\~\\~\\~}%leave this in to get the correct vertical space after the abstract
 
\end{abstract}

\begin{deluxetable}{llllll}
\tabletypesize{\small}
\tablecaption{Post-PPV rotational period distributions for young ($\leq$1~Gyr)
  stars \label{pref}}
\tablewidth{0pt}
\tablehead{Reference &  Target & Age & Mass range & N$_\star$ \\&& (Myr) & (M$_\odot$) }
\startdata
\citet{Grankin2013} & Taurus & 1-3 & 0.4-1.6 & 61 \\
\citet{Xiao2012} & Taurus & 1-3 & 0.3-1.2 & 18 \\
\citet{Artemenko2012} & Various SFRs & 1-5 & 0.3-3.0 & 52  \\
\citet{Henderson2012} & NGC 6530 & 2 & 0.2-2.0 &  244  \\
\citet{Rodriguez2009} & ONC & 2 & 0.015-0.5 & 487 \\
\citet{Affer2013} & NGC 2264 & 3 & 0.2-3.0 & 209 \\
\citet{Cody2010} & $\sigma$ Ori & 3 & 0.02-1.0  & 64 \\
\citet{Littlefair2010} & Cep OB3b & 4-5 & 0.1-1.3 & 475 \\
\citet{Irwin2008b} & NGC 2362 & 5 & 0.1-1.2 & 271 \\
\citet{Messina2011} & Young assoc. & 6-40 &  0.2-1.0 & 80 \\
\citet{Messina2010} & Young assoc. & 8-110 & 0.2-1.0 & 165 \\
\citet{Moraux2013} & h Per & 13 & 0.4-1.4 & 586 \\
\citet{Irwin2008a} & NGC 2547 & 40 & 0.1-0.9 & 176 \\
\citet{Scholz2009} & IC 4665 & 40 & 0.05-0.5 & 20 \\
\citet{Hartman2010} & Pleiades & 125 & 0.4-1.3 & 383  \\
\citet{Irwin2009} & M 50 & 130 & 0.2-1.1 & 812 \\
\citet{Irwin2007} & NGC 2516 &  150 &  0.15-0.7 & 362 \\
\citet{Meibom2009} & M35 & 150 & 0.6-1.6 & 310  \\
\citet{Sukhbold2009} & NGC 2301 & 210 & 0.5-1.0 & 133  \\
\citet{Meibom2011b} & M34 & 220 & 0.6-1.2 & 83  \\
\citet{Hartman2009} & M 37 & 550 & 0.2-1.3 & 371  \\
\citet{Scholz2007} & Praesepe & 578 & 0.1-0.5 & 5 \\
\citet{Agueros2011} & Praesepe & 578 & 0.27-0.74 & 40  \\
\citet{Scholz2011} & Praesepe & 578 & 0.16-0.42 & 26  \\
\citet{CollierCameron2009} & Coma Ber & 591 & FGK & 46  \\
\citet{Delorme2011} & Praesepe/Hyades & 578/625 & FGK & 52/70  \\
\citet{Meibom2011a} & NGC 6811 & 1000 & FGK & 71  \\
\citet{Irwin2011} & Field M dwarfs & 500-13000& 0.1-0.3& 41 \\
\citet{Kiraga2007} & Field M dwarfs & 3000-10000 & 0.1-0.7 &
31 \\
\enddata
\end{deluxetable}

\begin{figure*}
\includegraphics[scale=0.65,angle=-90]{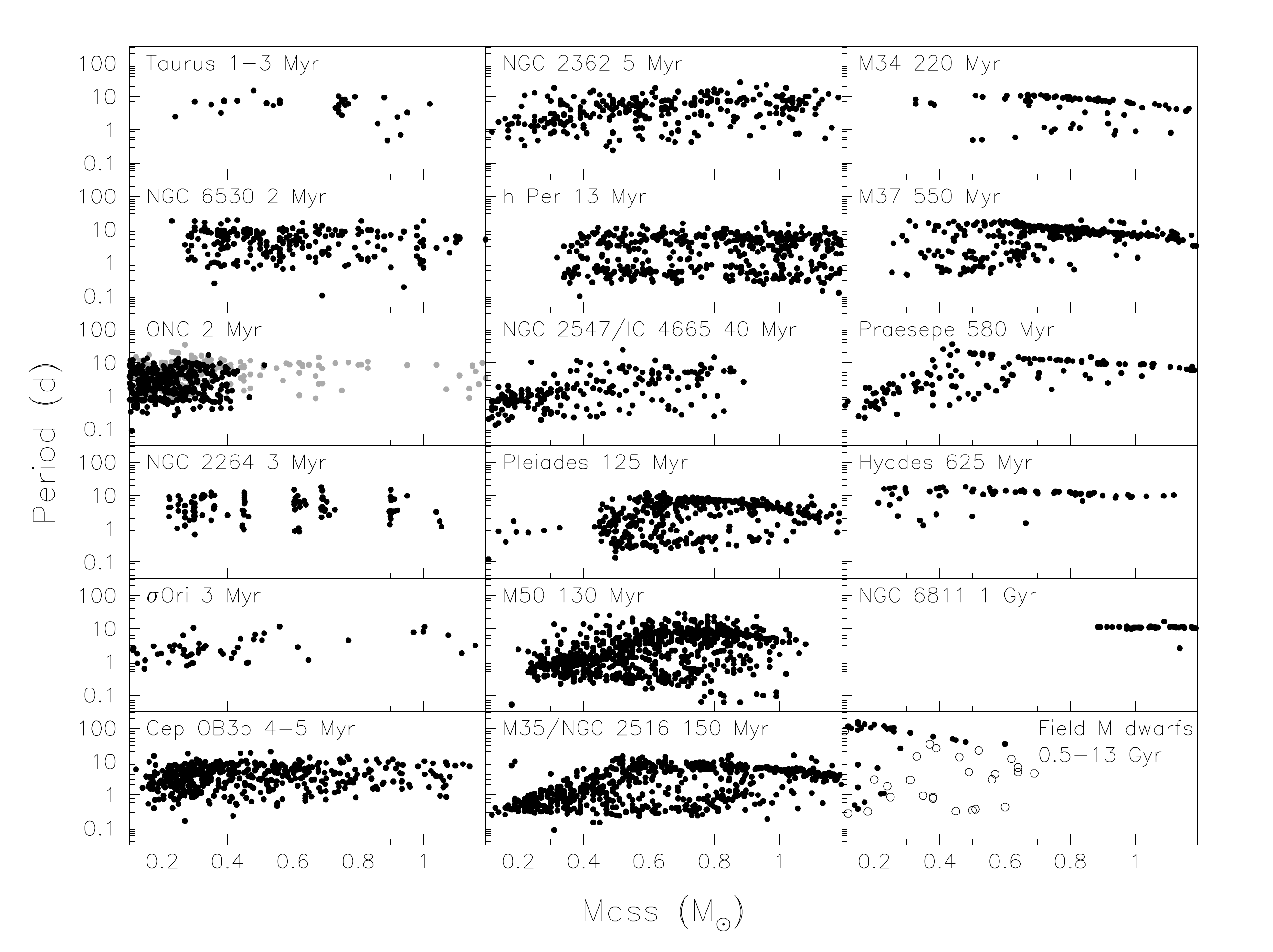}
\caption{\label{pdist}\small The rotational period distribution of low
  mass stars derived since Protostars \& Planets V in star forming
  regions, young open clusters, and in the field. The panels are
    ordered by increasing age, from top to bottom and left to right.
  The ONC panel includes previous measurements by \citet{Herbst2002}
  shown as grey dots. In the lower right panel, young disk M dwarfs
  are shown as open circles, old disk ones as filled
  circles. References are listed in Table~\ref{pref}.  }
 \end{figure*}

\bigskip
\centerline{\textbf{ 1. INTRODUCTION}}
\bigskip

The angular momentum content of a newly born star is one of the
fundamental quantities, like mass and metallicity, that durably
impacts on the star's properties and evolution. Rotation influences the
star's internal structure, energy transport, and the mixing processes in the stellar
interior that are reflected in surface elemental abundances. It is
also the main driver for magnetic activity, from X-ray luminosity to
UV flux and surface spots, that is the ultimate source of stellar
winds. Studying the initial angular momentum content of stars and its
evolution throughout the star's lifetime brings unique clues to the
star formation process, to the accretion/ejection phenomenon in young
stellar objects, to the history and future of stellar activity and its
impact on surrounding planets, and to physical processes that
redistribute angular momentum in stellar interiors.

Spectacular progress has been made, both on the observational and
theoretical sides, on the issue of the angular momentum evolution of
young stellar objects since Protostars \& Planets V. On the
observational side, thousands of new rotational periods have been
derived for stars over the entire mass range from solar-type stars
down to brown dwarfs at nearly all stages of evolution between birth
and maturity. The picture we have of the rotational evolution of
low-mass and very low-mass stars and brown dwarfs has never been as
well documented as of today. On the theoretical side, recent years
have seen a renaissance in numerical simulations of magnetized winds
that are the prime agent of angular momentum loss, new attempts have
been made to understand how young stars exchange angular momentum with
their disks via magnetic interactions, and new insights have been
gained on the way angular momentum is transported in stellar
interiors.

In the following sections, we review the latest developments which
shed new light on the processes governing the angular momentum
evolution of young stars and brown dwarfs, and also provide important
context for other Protostars \& Planets VI chapters to explore
possible connections between the rotational history of stars and the
formation, migration and evolution of planetary systems (star-disk
interaction, inner disk warps and cavities, planet engulfment,
irradiation of young planets, etc.). In Section 2, we review the
latest advances in the derivation of rotation rates for low mass stars
and brown dwarfs from birth to the early main sequence. In Section 3,
we provide an account of the physical mechanisms thought to dictate
the evolution of stellar rotation during the pre-main sequence (PMS)
and early main sequence (MS) , including star-disk interaction,
stellar winds, and angular momentum transport in stellar interiors. In
Section 4, we discuss various classes of angular momentum evolution
models that implement the latest theoretical developments to account
for the observed evolution of stellar rotation in cool stars and brown
dwarfs.

\bigskip
\textbf{ 2. OBSERVATIONAL STUDIES OF STELLAR ROTATION}
\bigskip

The measurement of rotational periods for thousands of stars in
molecular clouds and young open clusters provide the best way to trace
their angular momentum evolution from about 1~Myr to 1~Gyr. This
section discusses the observational studies of stellar rotation 
  performed since \citeapos{Herbst2007} PPV review, for solar-type stars and
lower mass stars, down to the brown dwarf regime.

\bigskip
\noindent
\textbf{ 2.1 Solar-mass and low-mass stars}
\bigskip

In the last 7 years, more than 5,000 new rotational periods have
  been measured for cool stars in star forming regions and young open
  clusters, over an age range from 1~Myr to 1~Gyr. In parallel,
  dedicated photometric monitoring of nearby M dwarfs, aimed at
  planetary transit searches, have reported tens of periods for the
  field very low-mass population over the age range 1-13~Gyr. These
  recent studies are listed in Table~\ref{pref} while Figure~\ref{pdist}
  provides a graphical summary of the results. A compilation of prior
  results was published by \citet{IrwinBouvier2009}.

  In addition, Kepler's and CoRoT's long term monitoring has provided
  rotation periods for more than 10,000 GKM field dwarfs
  \citep[e.g.,][]{Nielsen2013, McQuillan2013, Harrison2012,
    Affer2012}. These results offer a global view of stellar rotation
  as a function of mass on the main sequence, exhibiting a large
  dispersion at each spectral type, which possibly reflects the age
  distribution of the stellar samples.

The evolution of rotational distributions from 1~Myr to the old disk
population shown in Fig.~\ref{pdist} reveals a number of features:

\begin{itemize}
\item The initial distribution of spin rates at an age of about 2~Mr
  is quite wide over the whole mass range from 0.2 to 1.0~\msun, with
  the bulk of rotational periods ranging from 1 to 10 days. The lower
  envelope of the period distribution is located at about 0.7 days,
  which corresponds to about 40-50\% of the break-up limit over the
  mass range 0.2-1.0~\msun. The origin of
  the initial scatter of stellar angular momentum for low mass stars
  remains an open issue and probably reflects physical processes
  taking place during the embedded protostellar
  stage. \citet{Gallet2013} suggested that the dispersion of initial angular
  momenta may be linked to the protostellar disk mass.
\item From 1 to 5~Myr, i.e., during the early PMS evolution, the
  rotation rates of solar-type stars hardly evolve. In contrast, the
  lowest mass stars significantly spin up. \citet{Henderson2012}
  suggested that the increasing period-mass slope for lower mass stars
  can be used as an age proxy for very young
  clusters. \citet{Littlefair2010}, however, reported that the
  similarly aged (5~Myr) NGC~2362 and Cep~OB3b clusters exhibit quite
  a different rotational period distribution at low masses, which may
  point to the impact of environmental effects on rotation properties.
\item Past the end of the PMS accretion phase, the rotational
  distribution of the 13~Myr h~Per cluster members is remarkably flat
  over the 0.4-1.2~\msun\ range. The lower envelope of the period
  distribution, now located at about 0.2-0.3d, bears strong evidence for
  PMS spin up, as the freely evolving stars contract towards the
  ZAMS. In contrast, the slow rotators still retain periods close to
  8-10 days, a result interpreted as evidence for core-envelope
  decoupling in these stars \citep{Moraux2013}. Similar results are
  seen in the 40~Myr clusters IC~4665 and NGC~2547, with the addition
  of very low mass stars that are faster rotators and exhibit a steep
  period-mass relationship.
\item Once on the early MS (0.1-0.6~Gyr), a well-defined sequence of
  slow rotators starts to appear over the mass range 0.6-1.1~\msun\
  while the lower mass stars still retain fast rotation. This suggests
  a spin down timescale of order of a few 0.1~Gyr for solar-type
  stars, as angular momentum is carried away by magnetized winds. The
  development of a slow rotator sequence and its gradual evolution
  towards longer periods indeed serves as a basis to main sequence
  gyrochronology \citep{Barnes2007}.
\item By an age of 0.5-0.6~Gyr, all solar-type stars down to a mass of
  0.6~\msun\ have spun down, thus yielding a tight period-mass
  relationship, with the rotation rate decreasing towards lower masses
  \citep{Delorme2011}. In contrast, the very low mass stars still
  exhibit a large scatter in spin rates at that age
  \citep{Agueros2011}. It is only in the old disk population, by about
  10 Gyr, that the majority of lowest mass stars join the slow rotator
  sequence \citep{Irwin2011}. Clearly, the spin down timescale is a
  strong function of stellar mass, being much longer for the lowest
  mass stars than for solar-type ones \citep[e.g.,][]{McQuillan2013}.
\end{itemize} 

A long-standing and somewhat controversial issue remains the
so-called ``disk-locking'' process, i.e., the observational evidence
that stars magnetically interacting with their accretion disk during
the first few Myr of PMS evolution are prevented from spinning up in
spite of contracting towards the ZAMS
\citep[e.g.,][]{Rebull2004}. A number of post-PPV studies tend
to support the view that, at a given age, disk-bearing PMS stars are,
on average, slower rotators than diskless ones, with periods typically
in the range from 3 to 10 days for the former, and between 1 and 7
days for the latter \citep[e.g.,][]{Affer2013, Xiao2012,
  Henderson2012, Littlefair2010, Rodriguez2009, Irwin2008b,
  Cieza2007}. However, in all star forming regions investigated so
far, there is a significant overlap between the rotational
distributions of classical and weak-line T Tauri stars. Furthermore,
\citet{Cody2010} failed to find any evidence for a disk-rotation
connection among the very low mass members of the 3~Myr $\sigma$ Ori
cluster, which suggests it may only be valid over a restricted mass
range.

The lack of a clear relationship between rotation and disk accretion
may have various causes.  Observationally, the determination of
rotational period distributions relies on the assumption that the
photometric periods derived from monitoring studies arise from surface
spot modulation and therefore accurately reflect the star's rotational
period. Recently, \citet{Artemenko2012} questioned the validity of
this assumption for classical T Tauri stars. Based on the comparison
of photometric periods and $v\sin i$ measurements, they claimed that
in a fraction of classical T Tauri stars the measured periods
correspond to the Keplerian motion of obscuring circumstellar dust in
the disk and are significantly longer than the stellar rotational
periods \citep[see also][]{Percy2010}. \citet{Alencar2010}, however,
found that the photometric periods of classical T Tauri stars undergoing cyclical disk
obscuration were statistically similar to those of classical T Tauri stars dominated by
surface spots, thus suggesting that the obscuring dust is located
close to the co-rotation radius in the disk.

On the theorical side, the star-disk interaction may impact the star's
rotation rate in various ways, depending in particular on the ratio
between the disk truncation and co-rotation radii. \citet{LeBlanc2011}
have modeled the spectral energy distribution of young stars in IC 348
in an attempt to derive the disk inner radius and evaluate its
relationship with the star's rotational period. No clear trend emerges
from the ratio of inner disc radius to corotation radius when
comparing slow and fast rotators. It should be cautioned, however,
that SED modeling actually measures the inner {\it dusty} disk radius,
while the {\it gaseous} disc may extend further in
\citep[e.g.,][]{Carr2007}. Also, scattered light in the near-IR may
substantially alter the measurement of inner dust disk radius in T
Tauri stars \citep{Pinte2008}. 

\bigskip

\bigskip 

\bigskip 

\noindent
\textbf{ 2.2 Very low-mass stars and brown dwarfs}
\bigskip

Significant progress has recently been made in evaluating the
rotational properties of very low mass objects (VLM, masses below
$\sim 0.3\,M_{\odot}$), including brown dwarfs (BDs), i.e., objects with
masses too low to sustain stable hydrogen burning
($M<0.08\,M_{\odot}$). In the last Protostars and Planets review on
this subject \citep{Herbst2007}, about 200 periods for VLM objects in
the ONC and NGC2264, two 1-3\,Myr old star forming regions, were
discussed. In addition, smaller samples in other clusters were already
available at that time.  For brown dwarfs the total sample was limited
to about 30 periods, only a handful for ages $>$10~Myr, complemented
by $v\sin i$ measurements. We summarize in this Section the most
recent advances regarding the measurement of spin rates for very low
mass stars and brown dwarfs, and recapitulate the emerging picture for
the angular momentum evolution in the VLM domain.

 {\bf Star forming regions:} For the Orion Nebula Cluster, at an age
 of 1-2\,Myr, \citet{Rodriguez2009} have published several
 hundred new VLM periods. This includes more than 100 periods for
 brown dwarf candidates, the largest BD period sample in any region
 studied so far. A new period sample across the stellar/substellar
 regime in the slightly older $\sigma$\,Ori cluster has been presented
 by \citet{Cody2010}, including about 40 periods for VLM
 objects. In addition, the period sample from the {\it Monitor}
 project in the 4-5\,Myr old cluster NGC2362 contains about 20-30 periods
 in the VLM regime \citep{Irwin2008b} and the new period sample in
 the 4-5~Myr Cep OB3b region published by \citet{Littlefair2010}
 extends into the VLM regime. Taken together with the previously
 reported samples, there are now more than hundred VLM periods
 available at ages of 3-5\,Myr.

 From the period distributions in these very young regions, the
 following features are noteworthy:
\begin{itemize}

\item  VLM objects at young ages show a wide range of periods, similar to
more massive stars, from a few hours up to at least 2 weeks.

\item In all these samples there is a consistent trend of faster
  rotation towards lower masses. In the ONC, the median period drops
  from 5\,d for $M>0.4\,M_{\odot}$ to 2.6\,d for VLM stars and to 2\,d
  for BDs \citep{Rodriguez2009}. As noted by \citet{Cody2010} and
  earlier by \citet{Herbst2001}, this period-mass trend is consistent
  with specific angular momentum being only weakly dependent on mass
  below about 1~M$_\odot$. An intriguing case in the context of the
  period-mass relation is the Cep OB3b region
  \citep{Littlefair2010}. While the same trend is observed, it is much
  weaker than in the other regions. The VLM stars in Cep OB3b rotate
  more slowly than in other clusters with similar age. This may be a
  sign that rotational properties are linked to environmental factors,
  a possibility that needs further investigation.

\item A controversial aspect of the periods in star forming regions is
  their lower limit. The breakup limit, where centrifugal forces
  balance gravity, lies between 3 and 5\,h at these young
  ages. \citet{Zapatero2003}, \citet{Caballero2004}, and \citet{Scholz2004a,
    Scholz2005} report brown dwarf periods that are very close to that
  limit. On the other hand, the \citet{Cody2010} sample
  contains only one period shorter than 14\,h, although their
  sensitivity increases towards shorter periods. Thus, it remains to
  be confirmed whether some young brown dwarfs indeed rotate close to
  breakup speed.

\item Whether the disk-rotation connection observed for solar-mass and
  low-mass young stars extends down to the VLM and brown dwarf domains
  remains unclear.  \citet{Cody2010} found the same period
  distribution for disk-bearing and diskless VLM stars in the 3~Myr
  $\sigma$ Ori cluster while in the 2~Myr ONC \citet{Rodriguez2010}
  find that objects with NIR excess tend to rotate slower than objects
  without NIR excess in the mass regime between 0.075 and
  0.4$\,M_{\odot}$. No such signature is seen in the substellar regime
  with the possible caveat that many brown dwarf disks show little or
  no excess emission in the NIR and require MIR data to be clearly
  detected. Finally, \citet{Mohanty2005}, \citet{Nguyen2009},
  \citet{Biazzo2009}, and \citet{Dahm2012} report somewhat conflicting
  results regarding the existence of a disk-rotation connection in the
  very low mass regime based on $v \sin i$ measurements of members of
  1-5~Myr clusters .

\end{itemize}

{\bf Pre-main sequence clusters:} For the pre-main sequence age range
between 5 and 200\,Myr, about 200 VLM periods are now available, a
factor of 20 increase compared with 2007. About 80 of them have been
measured in IC4665 and NGC2547, two clusters with ages around 40 Myr
\citep{Scholz2009, Irwin2008a}. Approximately 100 periods are
available for VLM objects in NGC2516 ($\simeq$150~Myr) from
\citet{Irwin2007}. In addition, a few more VLM periods are contained
in the samples for M34 \citep{Irwin2006} and M50 \citep{Irwin2009},
although the latter sample might be affected by substantial
contamination. Note that in these clusters (as well as in the
main-sequence Praesepe cluster discussed below) the number of measured
BD periods is very low (probably $<5$).

The most significant feature in the period distributions in this mass
and age regime is the distinctive lack of slow rotators. Essentially
all VLM periods measured thus far in these clusters are shorter than
2\,d, with a clear preference for periods less than 1\,d. The median
period is 0.5-0.7\,d. The lowest period limit is around 3\,h. This
preference for fast rotators cannot be attributed to a bias in the
period data, for two reasons. First, most of the studies cited above
are sensitive to longer periods. Second, \citet{Scholz2004b}
demonstrate that the lower envelope of the $v\sin i$ for Pleiades VLM
stars \citep{Terndrup2000} translates into an upper period limit of
only 1-2\,d, consistent with the period data.

Compared with the star forming regions, both the upper period limit
and the median period drop significantly. As discussed by
\citet{Scholz2009}, this evolution is consistent with angular momentum
conservation plus weak rotational braking, but cannot be accomodated
by a Skumanich-type wind braking law.

{\bf Main sequence clusters:} For the Praesepe cluster, with an age of
580\,Myr a cornerstone for tracing the main-sequence evolution of
stars, around 30 rotation periods have been measured for VLM stars
\citep{Scholz2007, Scholz2011}. In combination with the samples for
more massive stars by \citet{Delorme2011} and \citet{Agueros2011},
this cluster has now a well defined period sample for FGKM dwarfs
(cf. Fig.~\ref{pdist}). With one exception, all VLM stars in the
Scholz et al. sample have periods less than 2.5\,d, with a median
around 1\,d and a lower limit of 5\,h. \citet{Scholz2011} compared
the Praesepe sample with the pre-main sequence clusters. While the
evolution of the lower period limit is consistent with zero or little
angular momentum losses between 100 and 600\,Myr, the evolution of the
upper period limit implies significant rotational braking. In this
paper, this is discussed in terms of a mass-dependent spindown on the
main sequence. With an exponential spindown law, the braking timescale
$\tau$ is $\sim$0.5~Gyr for 0.3$\,M_{\odot}$, but $>1$\,Gyr for
0.1$\,M_{\odot}$. Thus, wind braking becomes less efficient towards
lower masses. Similar to the pre-main sequence clusters, the rotation
of brown dwarfs is unexplored in this age regime.

{\bf Field populations:} The largest (in fact, the only large) sample
of periods for VLM stars in the field has been published recently by
\citet{Irwin2011}, with 41 periods for stars with masses between the
hydrogen burning limit and 0.35$\,M_{\odot}$. A few more periods in
this mass domain are available from \citet{Kiraga2007}. Interestingly,
the Irwin et al. sample shows a wide spread of periods, from 0.28\,d
up to 154\,d. in stark contrast to the uniformly fast rotation in
younger groups of objects. Based on kinematical age estimates, Irwin
et al. find that the majority of the oldest objects in the sample
(thick disk, halo) are slow rotators, with a median period of 92\,d. For
comparison, the younger thin disk objects have a median period of
0.7\,d. This provides a firm constraint on the spindown timescale of
VLM stars, which should be comparable with the thick disk age, i.e.,
8-10\,Gyr. Similar conclusions were reached by \citet{Delfosse1998}
and \citet{Mohanty2003} based on $v\sin i$ measurements.

Brown dwarfs in the field have spectral types of L, T, and Y, and
effective temperatures below 2500\,K. At these temperatures, magnetic
activity as it is known for M dwarfs, is not observed anymore, thus,
periodic flux modulations from magnetically induced spots as in VLM
stars are not expected. Some L- and T-dwarfs, however, do exhibit
persistent periodic variability, which is usually attributed to the
presence of a non uniform distribution of atmospheric clouds
\citep[see][]{Radigan2012}, which again allows for a measurement of
the rotation period. About a dozen of periods for field L- and
T-dwarfs are reported in the literature and are most likely the
rotation periods \citep{Bailer-Jones1999, Bailer-Jones2001,
  Clarke2002, Koen2006, Lane2007, Artigau2009,
  Radigan2012, Heinze2013}. All these periods are shorter than 10\,h.

Again it is useful to compare these findings with $v\sin i$ 
data. Rotational velocities have been measured for about 100 brown
dwarfs by \citet{Mohanty2003}, \citet{Zapatero2006},
\citet{Reiners2008, Reiners2010}, \citet{Blake2010}, and
\citet{Konopacky2012}. The composite figure by \citet[][their
Fig.3]{Konopacky2012} contains about 90 values for L- and 8 for
T-dwarfs. From this combined dataset it is clear that essentially all
field brown dwarfs are fast rotators with $v\sin i >$7 km s$^{-1}$,
corresponding to periods shorter than 17\,h, thus confirming the
evidence from the smaller sample of periods. The lower limit in $v\sin
i$ increases towards later spectral types, from 7~km s$^{-1}$ for
early L dwarfs to more than 20~km s$^{-1}$ for late L dwarfs and
beyond. Because brown dwarfs cool down as they age and thus spectral
type is a function of age and mass, these values are difficult to
compare with models. However, they strongly suggest that rotational
braking becomes extremely inefficient in the substellar
domain. Extrapolating from the trend seen in the M dwarfs, the braking
timescales for brown dwarfs are expected to be longer than the age of
the Universe.

The results of rotational studies of the VLM stars in Praesepe and in
the field clearly indicate that the spindown timescale increases
towards lower masses in the VLM regime. There is no clear mass
threshold at which the long-term rotational braking ceases to be
efficient, as might be expected in a scenario where the dynamo mode
switches due to a change in interior structure. The observational data
rather suggests that the rotational braking becomes gradually less
efficient towards lower masses, until it essentially shuts down for
brown dwarfs.

\bigskip
\noindent
\textbf{ 2.3 Gyrochronology}
\bigskip

By the time low-mass stars reach the ZAMS (100--200 Myr), the
observations show clear evidence for two distinct sequences of fast
and slow rotators in the mass vs.\ period plane, presumably tracing
the lower and upper envelopes of stellar rotation periods at the
ZAMS. Observations in yet older open clusters show a clear convergence
in the angular momentum evolution for all FGK dwarfs towards a single,
well-defined, and mass-dependent rotation period by the age of the
Hyades ($\sim$600 Myr, cf. Fig~\ref{pdist}).

These patterns---and in particular the observed sequence of slow
rotators with stellar mass---have been used as the basis for
``gyrochronology" as an empirical tool with which to measure the ages
of main-sequence stars \citep[e.g.,][]{Barnes2003, Barnes2007,
  Mamajek2008, Meibom2009, CollierCameron2009, Delorme2011}.  The
method has so far been demonstrated and calibrated for solar-type
stars with convective envelopes (i.e., mid-F to early-M dwarfs), with
ages from the ZAMS to the old field population.

In the gyrochronology paradigm, the principal assumptions are that a
given star begins its main-sequence rotational evolution with a
certain ``initial" rotation period on the ZAMS, on either a
rapid-rotator sequence (so-called `C' sequence) or a slow-rotator
sequence (so-called `I' sequence).  All stars on the rapid-rotator
sequence evolve onto the slow-rotator sequence on a timescale governed
by the convective turnover time of the convective envelope
\citep{Barnes2010a}.  Once on the slow-rotator sequence, the star then
spins down in a predictable way with time, thus allowing its age to be
inferred from its rotation period.

As discussed by \citet{Epstein2012}, there are limitations to the
technique, particularly in the context of very young stars. First, to
convert a given star's observed rotation period into a gyro-age
requires assuming the star's initial rotation period. At older ages,
this is not too problematic because the convergence of the
gyro-chrones makes the star eventually forget its own initial ZAMS
rotation period. However, at ages near the ZAMS, the effect of the
(generally unknown) initial period is more important.

More importantly, the technique has not yet been demonstrated or
calibrated at ages earlier than the ZAMS.  Presumably, the rotational
scatter observed at the ZAMS and the relationships between stellar
mass and surface rotation period must develop at some stage during the
PMS and should be understood in the context
of the early angular momentum evolution of individual stars (see
Section 4). Interestingly, there is now observational evidence that
specific patterns may be emerging during the PMS in the period-mass
diagrams, which encode the age of the youngest clusters \citep{Henderson2012}.

\bigskip
\textbf{ 3. THE PHYSICAL PROCESSES GOVERNING ANGULAR
MOMENTUM EVOLUTION}
\bigskip

The evolution of the rotation period and the angular momentum of a
forming protostar is determined both by internal and external
processes.  External processes include all the mechanisms of angular
momentum exchange with the surrounding ambient medium, with the
accretion disk and the circumstellar outflows in particular. Internal
processes determine the angular momentum redistribution throughout the
stellar interior. These various processes are discussed in this
section.

\begin{figure}
\includegraphics{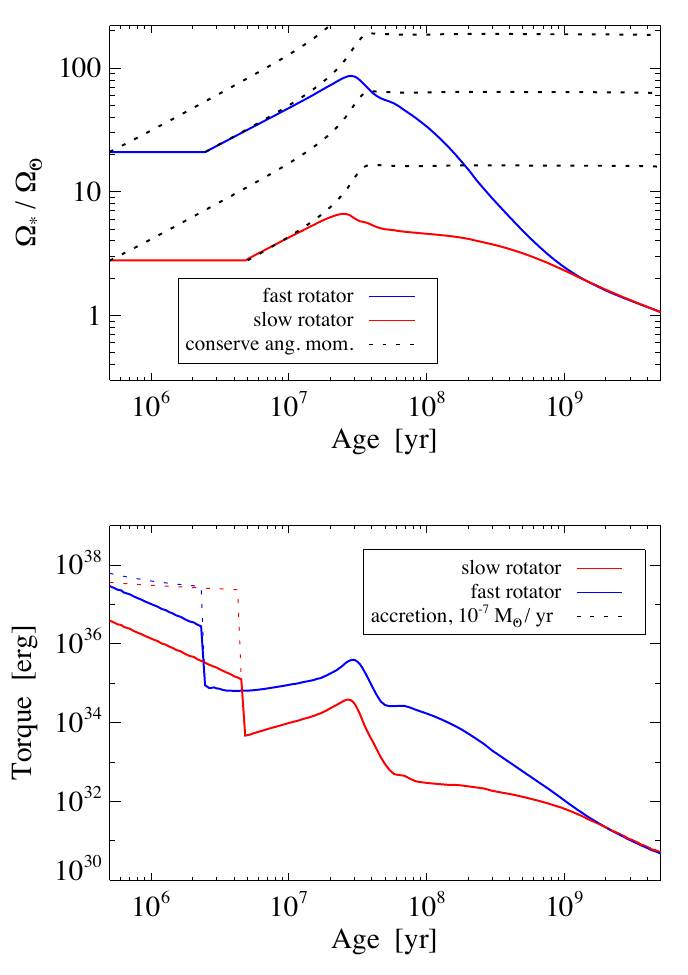}
\caption{ \small {\it Upper panel:} Angular rotation rate of a 1
    $M_\odot$ star as a function of age. The red and blue tracks show
    the evolution of the surface rotation rate from the
    \citeapos{Gallet2013} models, which best reproduce the lower and
    upper (respectively) range of the observed spin distributions
    (cf. Fig.~\ref{gb13}) . The dotted lines show the expected
    evolution of rotation rate, if the star were to conserve angular
    momentum, shown at a few different ``starting points.'' {\it Lower
      panel:} The external torque on the star that is required to
    produce the spin evolution tracks of the upper panel.  The red and
    blue lines show the torque required, assuming only the stellar
    structural evolution from a model of \citet{Baraffe:1998p5181} and
    the (solid-body) spin evolution of the corresponding red and blue
    lines in the upper panel.  The dotted lines show the torque that
    would be required if the stars were also accreting at a rate of
    $10^{-7} M_\odot yr^{-1}$, during the time when the rotation rate
    is constant on the upper panel tracks. 
\label{fig_hypotorque}}
 \end{figure}

    The upper panel of Figure \ref{fig_hypotorque} shows the
      evolution of the surface rotation rate of a 1~M$_\odot$ star, as
      predicted by \citeapos{Gallet2013} models.  The blue and red tracks
      correspond to the upper and lower envelopes of the observed spin
      distributions, and represent the range of spin evolutions
      exhibited by the majority of solar-mass stars
      (cf. Fig.~\ref{gb13}).  The dotted lines show the evolution of
      the rotation rate, assuming that the star conserves angular
      momentum (assuming solid body rotation and structural evolution
      from \citealp{Baraffe:1998p5181}).  These angular momentum
      conserved tracks are shown for a few arbitrary ``starting
      points,'' at the earliest time shown in the plot and at the time
      where the rotation rate is no longer constant in time.  Assuming
      angular momentum conservation, the stars are expected to spin
      up, due to their contraction as they evolve toward the zero-age
      main sequence (at $\sim$40~Myr) and then to have a nearly
      constant rotation rate on the main sequence (since then the
      structure changes much more slowly).  It is clear from a
      comparision between the dotted lines and solid lines that the
      observed evolution is completely inconsistent with angular
      momentum conservation and instead requires substantial angular
      momentum loss.

      The lower panel of Figure \ref{fig_hypotorque} shows the
      external torque on the star that is necessary to produce the red
      and blue tracks of the upper panel, assuming solid body
      rotation.  If the star is also accreting from a Keplerian disk,
      this should in principle result in an additional spin-up torque
      on the star, given approximately by $\tau_{\rm a} \ga \dot
      M_{\rm a} (G M_* R_*)^{1/2}$, where $\dot M_{\rm a}$ is the
      accretion rate \citep[e.g.,][]{ghoshlamb78, Matt2005b}.  As a
      simple quantitative example, the dotted lines in the lower panel
      show the torque required to produce the gyrotracks of the upper
      panel, assuming the stars are accreting at a constant rate of
      $\dot M_{\rm a} = 10^{-7} M_\odot yr^{-1}$, during the time when
      the rotation rate is constant.

      It is clear from the figure that the observed evolution of the
      spin rates of solar-mass stars requires substantial angular
      momentum loss at nearly all stages, and the required torque is
      largest at the youngest ages.  During the first few Myr of the T
      Tauri phase, the observed spin-rate distributions do not appear
      to evolve substantially.  The torques required simply to prevent
      these stars from spinning up, due to contraction, are
      $\sim$10$^6$ times larger than the torque on the present day
      Sun.  Accreting stars require even larger torques to counteract
      the additional spin-up effects of accretion, which depends
      primarily on the accretion rate.

 After an age of $\sim$5~Myr, and until the stars reach the ZAMS (at
 $\sim$40~Myr), the stars spin up, due to their contraction.  Although
 the surface rotation rates suggest a substantial fraction of their
 angular momentum is lost during this spin-up phase, the torque is
 apparently much less than during the first few Myr. This apparent
 sudden change in the torque suggests a change in the mechanism
 responsible for angular momentum loss.  The fact that a substantial
 fraction of stars younger than $\sim$5~Myr are still accreting
 suggests that the star-disk interaction may in some way be
 responsible for the largest torques \citep{Koenigl1991}.  In this case,
 the eventual dissipation of the disk (i.e., the cessation of
 accretion) provides a natural explanation (at least in principle) for
 the transition to much weaker torques.

\newpage
\bigskip
\noindent
\textbf{3.1 Star-disk interaction}
\bigskip

\begin{figure}[t]
\includegraphics[width=\linewidth]{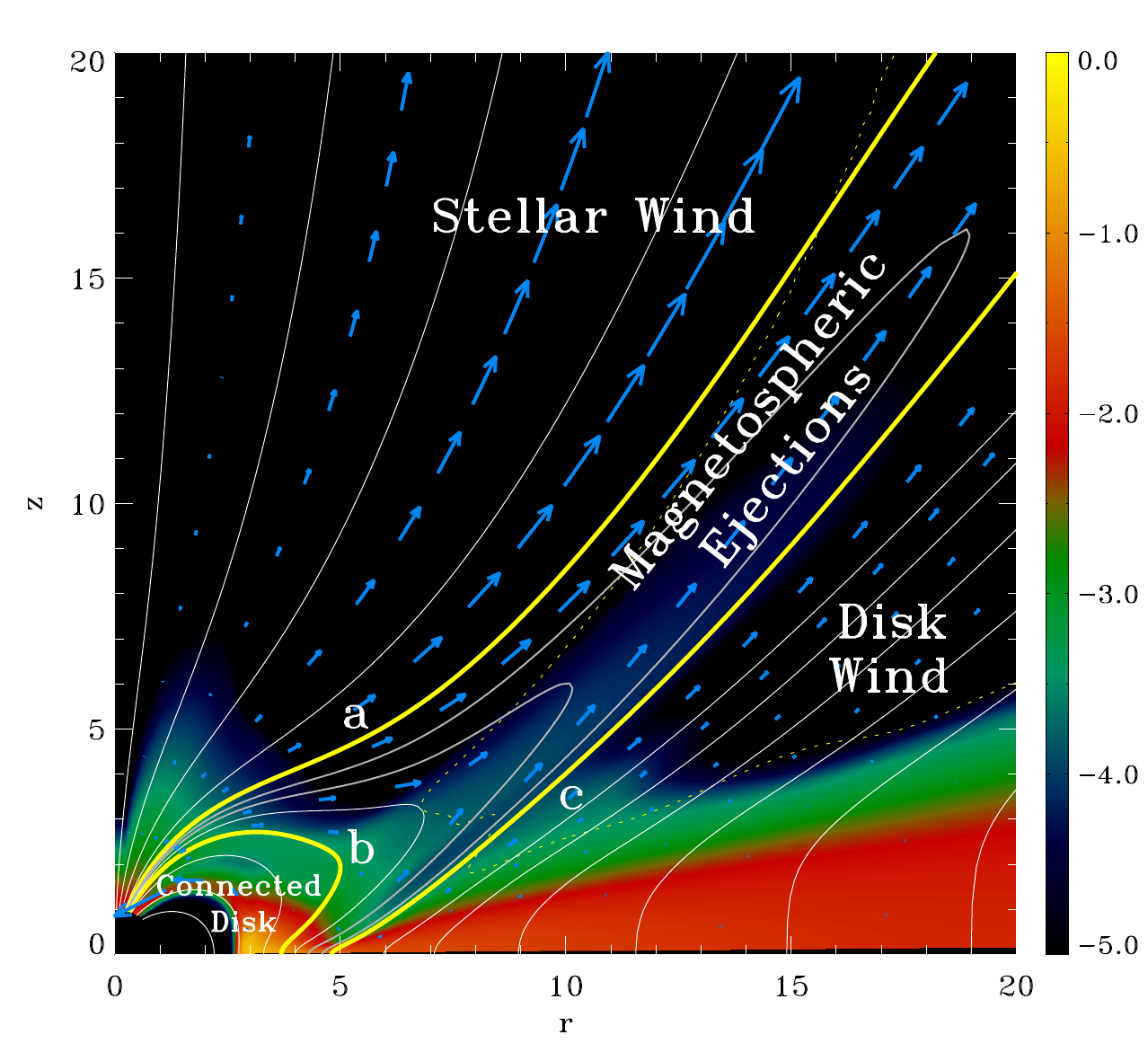}
\caption{\label{sdi} \small Schematic view of the outflows that can be found in a
  star-disk interacting system. 1) Stellar winds accelerated along the
  open magnetic flux anchored onto the star; 2) magnetospheric
  ejections associated with the expansion and reconnection processes of
  closed magnetic field lines connecting the star and the disk; 3)
  disk-winds (including X-winds) launched along the open stellar
  magnetic surfaces threading the disk. From \citet{Zanni2013}.}
\end{figure}

Our understanding of the various processes that are involved in the
magnetic interaction between the star and its surrounding accretion
disk has improved significantly in the last few years
\citep[e.g.,][]{Bouvier2007}.  A number of recent results have spurred
the development of new models and ideas for angular momentum
transport, as well as further development and modification to existing
models. In particular, it has been clear for some time that, due to
the differential twisting of the magnetic field lines, the stellar
magnetic field cannot connect to a very large region in the disk
\citep[e.g.,][]{Shu1994, Lovelace1995, Uzdensky2002, Matt2005b}. In
addition, the competition between accretion and diffusion is likely to
reduce the magnetic field intensity in the region beyond the
corotation radius \citep{Bardou1996, Agapitou2000, Zanni2009}. As a
consequence, the widespread ``disk-locking'' paradigm, as proposed in
the classical Ghosh \& Lamb picture \citep{Ghosh1979}, has been
critically re-examined.  Also, it has been realized that the strength
of the dipolar components of magnetic fields are generally
significantly weaker than the average surface fields, which indicates
that the latter is dominated by higher order multipoles
\citep{Gregory2012}. As the large-scale dipolar field is thought to be
the key component for angular momentum loss, mechanisms for extracting
angular momentum from the central star are thus required to be even
more efficient.  These issues have prompted different groups to
reconsider and improve various scenarios based on the presence of
outflows that could efficiently extract angular momentum from the
star-disk system, as schematically illustrated in Figure~\ref{sdi}.
The key new developments have been (1) the generalization of the
X-wind model to multi-polar fields, (2) a renewed exploration of
stellar winds as a significant angular momentum loss mechanism, and
(3) the recognition that magnetospheric ejections that naturally arise
from the star-disk interaction may actually have a significant
contribution to the angular momentum transport.

Note that in the following sections we deal with only one specific
class of disk winds, namely the X-wind, and neglect other popular
models, the "extended disk wind" in particular, presented in {\it
  A. Frank et al.}'s chapter. First, in its "standard" formulation
\citep[e.g.,][]{Blandford:1982p3911, Ferreira1997}, an extended disk
wind exerts a torque onto the disk without modifying its Keplerian
angular momentum distribution. In this respect, such a solution simply
allows the disk to accrete and it does not affect the stellar angular
momentum evolution. Second, we only discuss models that exploit the
stellar magnetic flux. It has been shown \citep[e.g.,][]{Zanni2013}
that the stellar magnetic field cannot provide enough open flux to the
disk to produce a relevant extended disk wind: the latter needs a
proper disk field distribution to be powered. Scenarios in which a
disk field interacts with the stellar magnetic flux have been proposed
\citep[see][]{Ferreira:2000p1706} and they deserve future
investigation.

\begin{figure*}[ht]
\includegraphics[width=\linewidth]{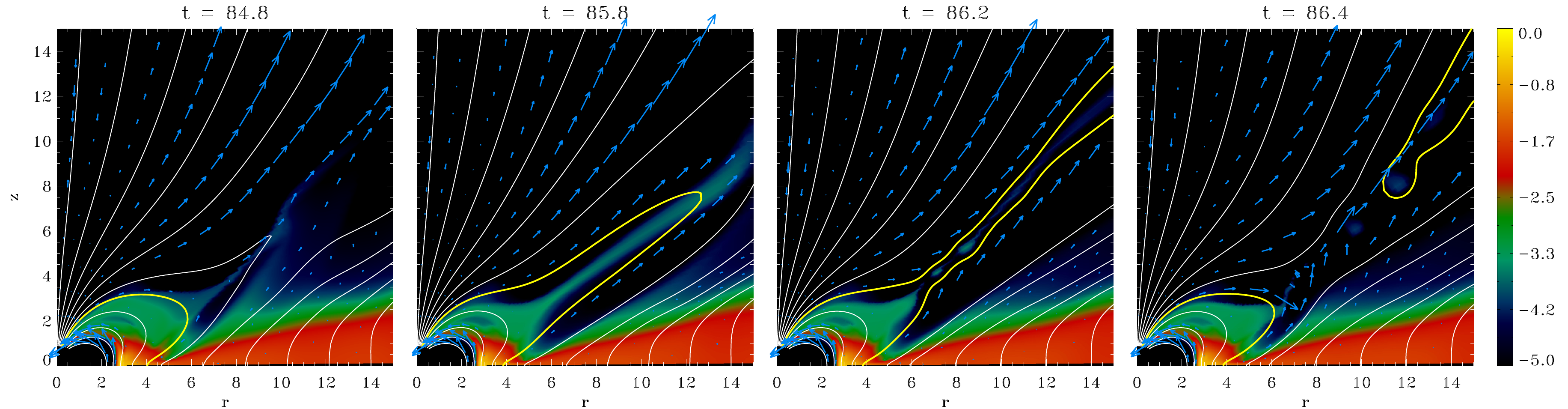}
\caption{\label{cycle} \small Temporal evolution of the periodic
  inflation/reconnection process associated with magnetospheric
  ejections. Speed vectors (blue arrows) and magnetic field lines
  (white lines) are superimposed to logarithmic density maps. The
  solid yellow line follows the evolution of a single magnetic
  surface. Time is given in units of the stellar rotation period. From
  \citet{Zanni2013}.}
\end{figure*}

\bigskip
\noindent
\textbf{ 3.1.1 Accretion driven stellar winds}
\bigskip

The idea that stellar winds may be important for removing angular
momentum from accreting stars has been around since the first
measurements of the rotational properties of young stars
\citep[e.g.,][]{Shu1988, Hartmann1989}.  For stars that
  are actively accreting from a disk, the amount of angular momentum
  carried onto the star by the disk is proportional to the accretion
  rate.  Stellar winds could be important for counteracting the
  spin-up effect due to accretion, if the mass outflow rate is a
  significant fraction ($\sim 10$\%) of the accretion rate
  \citep{Hartmann1989, Matt2005a}.  Due the substantial
  energy requirements for driving such a wind, \citet{Matt2005a}
  suggested that a fraction of the gravitational potential energy
  released by accretion (the ``accretion power'') may power a stellar
  wind with sufficiently enhanced mass outflow rates.

The torque from a stellar wind (discussed further below) depends upon
many factors, and generally increases with magnetic field strength and
also with mass loss rate.  Thus, the problem discussed above of having
weak dipolar fields can in principle be made up for by having a larger
mass loss rate.  However, as the mass loss rate approaches a
substantial fraction of the mass accretion rate, there will not be
enough accretion power to drive the wind. \citet{Matt2008} 
derived a ``hard'' upper limit of $\dot{M}_{wind}/\dot{M}_{acc}\la
60$\%.  By considering that the accretion power must be shared between
(at least) the stellar wind and the observed accretion diagnostics
(e.g., the UV excess luminosity), \citet{Zanni2011} made a
quantitative comparison between the predictions of accretion-powered
stellar winds and an observed sample of accreting stars, to test
whether there was enough accretion power available to drive a wind
capable of removing the accreted angular momentum.  The range of
observational uncertainties in quantities such as the UV excess and
dipolar magnetic field strength was sufficient to span the range of
possibilities from having enough accretion power to not having enough
accretion power.  However, this work demonstrated again that
accretion-powered stellar winds need substantial large-scale magnetic
fields in order to be efficient, and the strength of the observed
fields are (within uncertainties) near the mimimun required strengths
\citep{Gregory2012}.

Even if there is enough accretion power to drive a wind, there is
still a question of whether or how accretion power may transfer to the
open field regions of a star and drive a wind.  Based on calculations
of the emission properties and cooling times of the gas,
\citet{Matt2007} ruled out solar-like, hot coronal winds for mass loss
rates greater than $\sim 10^{-11}$ M$_\odot$/yr for ``typical'' T
Tauri stars of $\sim 0.5$ M$_\odot$.  They concluded that more massive
winds would be colder (colder than $\sim 10,000$K) and thus must be
driven by something other than thermal pressure, such as Alfv\'en
waves \citep{Decampli1981}.  Taking a ``first principles'' approach
and adopting a simplified, 1-D approach, \citet{Cranmer2008,
  Cranmer2009} developed models whereby the energy associated with
variable accretion drove MHD waves that traveled from the region of
accretion on the star to the polar regions where it led to enhanced
MHD wave activity and drove stellar wind.  Those models typically
reached mass loss rates that were equal or less than a few percent of
the accretion rate.  These values appear to be on the low end of what
is needed for significant angular momentum loss for most observed
systems.  Further theoretical work is needed to explore how accretion
power may transfer to a stellar wind.

\bigskip
\bigskip
\bigskip
\noindent
\textbf{ 3.1.2 Magnetospheric ejections}
\bigskip

Magnetospheric ejections (MEs) are expected to arise because of the
expansion and subsequent reconnection of the closed magnetospheric
field lines connecting the star to the disk \citep{Hayashi1996,
  Goodson1999, Zanni2013}.  The inflation process is the result of the
star-disk differential rotation and the consequent build-up of
toroidal magnetic field pressure.  This is the same phenomenon that
bounds the size of the closed magnetosphere and limits the efficiency
of the \citet{Ghosh1979} mechanism.  Initially, MEs are launched along
magnetic field lines which still connect the star with the disk so
that they can exchange mass, energy and angular momentum both with the
star and the disk. On a larger spatial scale, the MEs disconnect from
the central region of the disk-star system in a magnetic reconnection
event and propagate ballistically as magnetized plasmoids.  Because of
magnetic reconnection, the inner magnetic surfaces close again and the
process repeats periodically (see Fig. \ref{cycle}). This phenomenon
is likely to be related to the ``conical winds'' simulated by
\citet{Romanova2009}.

MEs contribute to control the stellar rotation period in two ways
\citep{Zanni2013}.  First, they extract angular momentum
from the disk close to the truncation region so that the spin-up
accretion torque is sensibly reduced.  This effect closely resembles
the action of an X-wind (see next subsection), which represents the limit
solution capable of extracting all the angular momentum carried by the
accretion flow. Second, if the ejected plasma rotates slower than the
star, the MEs can extract angular momentum directly from the star
thanks to a differential rotation effect. In such a situation, MEs are
powered by both the stellar and the disk rotation, as in a huge
magnetic slingshot. In agreement with other popular scenarios
\citep{Koenigl1991, Ostriker1995}, the spin-down torque
exerted by the MEs can balance the accretion torque when the disk is
truncated close to the corotation radius. In a propeller phase
\citep[see, e.g.,][]{Ustyugova2006,Dangelo2011}, when the truncation radius gets
even closer to corotation, the spin-down torque can even balance the
spin-up due to contraction.

Despite these first encouraging results, various issues remain. The
MEs phenomenon is based on a charge and discharge process whose
periodicity and efficiency depend on magnetic reconnection events that
are controlled by numerical diffusion only in the solutions proposed
by \citet{Zanni2013}. In order to produce an efficient spin-down
torque, this scenario requires a rather strong kG dipolar field
component, which has been only occasionally observed in classical T
Tauri stars \citep[e.g.,][]{Donati2008, Donati2010b}. In the propeller
regime, which provides the most efficient spin-down torque, the
accretion rate becomes intermittent on a dynamical timescale,
corresponding to a few rotation periods of the star. Even though this
effect is enhanced by the axial symmetry of the models, there is as
yet no observational evidence for such a behavior.

\bigskip
\noindent
\textbf{ 3.1.3 X-Winds}
\bigskip

\begin{figure}[t]
\includegraphics[width=\linewidth]{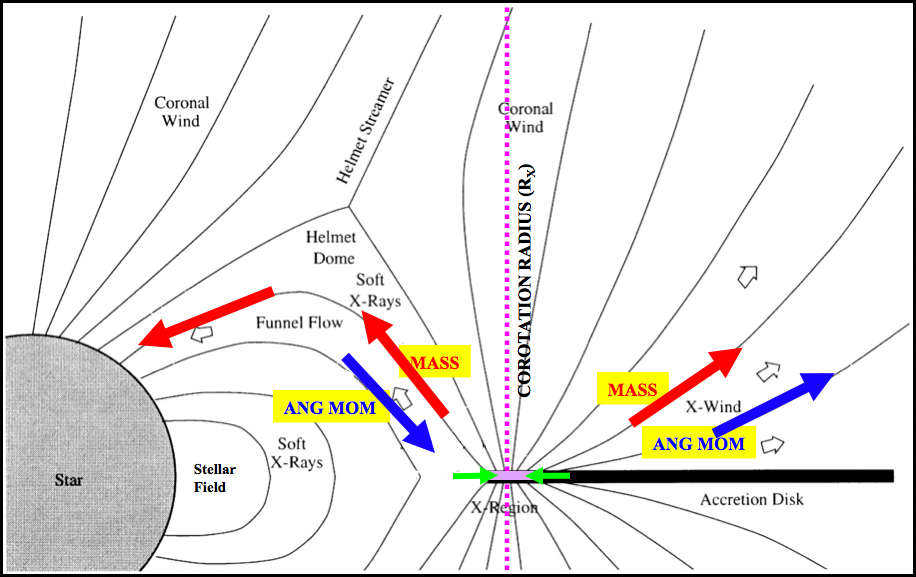}
\caption{\label{fig-xwind_schematic} Schematic of steady-state
    X-wind model. {\it Black thick line} in the equatorial plane is
    the truncated disk; {\it black solid curves} show the magnetic
    field; {purple dotted line} shows the co-rotation radius $R_X$;
    {\it purple thick horizontal line} shows the X-region. {Red and
      blue arrows} show the direction of mass and angular momentum
    transport respectively: interior to $R_X$, material flows from the
    X-region onto the star in a funnel flow along field lines that bow
    sufficiently inwards, and the excess angular momentum in this gas
    flows back into the X-region via magnetic stresses; exterior to
    $R_X$, material flows out of the X-region in a wind, along field
    lines that bow sufficiently outwards, and carries away with it
    angular momentum from the X-region. {\it Green arrows} show the
    pinching of gas in the X-region due to the angular momentum
    transport, which truncates the disk at the inner edge and keeps
    magnetic flux trapped in the X-region.}
\end{figure}

\begin{figure*}[t]
\center
\includegraphics[width=0.7\linewidth,angle=270]{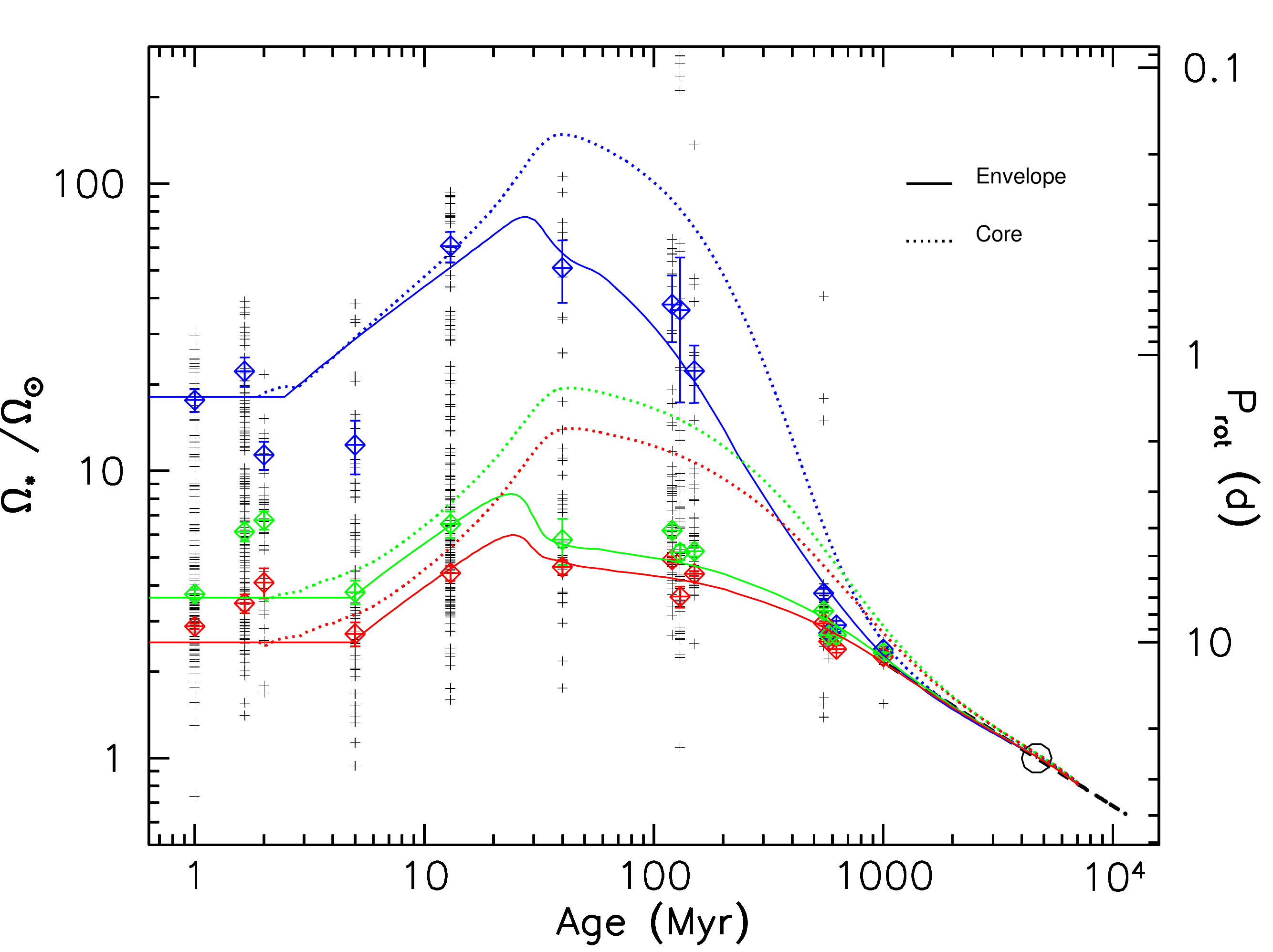}
\caption{\label{gb13} The rotational angular velocity of solar-type
  stars is plotted as a function of age. The left y-axis is labelled
  with angular velocity scaled to the angular velocity of the present
  Sun while the right y-axis is labelled with rotational period in
  days. On the x-axis the age is given in Myr. {\it Observations:} The
  black crosses shown at various age steps are the rotational periods
  measured for solar-type stars in star forming regions and young open
  clusters over the age range 1 Myr-1 Gyr. The red, green,
  and blue diamonds represent the 25, 50, and 90th percentiles of the
  observed rotational distributions, respectively. The open circle at 4.56 Gyr is
  the angular velocity of the present Sun. {\it Models:} The angular
  velocity of the convective envelope (solid line) and of the
  radiative core (dashed lines) is shown as a function of time for
  slow (red), median (green), and fast (blue) rotator models, with
  initial periods of 10, 7, and 1.4 days, respectively. The dashed
  black line at the age of the Sun illustrates the asymptotic
  Skumanich relationship, $\Omega\propto t^{ -1/2}$. From \citet{Gallet2013}.}
\end{figure*}

The X-wind model\footnote{Other types of outflows are considered
    in {\it Frank et al.}'s chapter} invokes the interaction of the stellar
magnetosphere with the surrounding disk to explain the slow spin rates
of accreting T Tauri stars, well below break-up, within a single
theoretical framework, via the central concept of {\it trapped flux}.
In steady-state, the basic picture is as follows (see
Fig.\,\ref{fig-xwind_schematic}): all the stellar magnetic flux
initially threading the entire disk is trapped within a narrow annulus
(the X-region) at the disk inner edge. The X-region straddles the
corotation radius $R_X$ (where the disk Keplerian angular velocity,
$\Omega_X = \sqrt{G M_{\ast}/R_X^3}$, equals the stellar angular
velocity, $\Omega_{\ast}$; $R_X$ lies near, but exterior to, the
inner-edge), a feature known as disk-locking. The resulting dominance
of the magnetic pressure over gas within the X-region makes the entire
annulus rotate as a solid body at the corotation angular velocity
$\Omega_X = \Omega_{\ast}$. Consequently, disk material slightly
interior to $R_X$ rotates at sub-Keplerian velocities, allowing it to
climb efficiently onto field lines that bow sufficiently inwards and
accrete onto the star; conversely, material within the X-region but
slightly exterior to $R_X$ rotates at super-Keplerian velocities,
enabling it to ascend field lines that bend sufficiently outwards and
escape in a wind.  The magnetic torques associated with the accretion
funnels transfer excess specific angular momentum (excess relative to
the amount already residing on the star) from the infalling gas to the
disk material at the footpoints of the funnel flow field lines in the
inner parts of the X-region, which tends to push this material
outwards. Conversely, the magnetic torques in the wind cause the
outflowing gas to gain angular momentum at the expense of the disk
material connected to it by field lines rooted in the outer parts of
the X-region, pushing this material inwards. The pinch due to this
outward push on the inside, and inward push on the outside, of the
X-region is what keeps the flux trapped within it, and truncates the
disk at the inner-edge in the first place. The net result is a
transfer of angular momentum from the accreting gas to the wind,
allowing the star to remain slowly rotating.

The X-wind accretion model was originally formulated assuming a dipole
stellar field \citep{Ostriker1995}. However, detailed
spectropolarimetric reconstructions of the stellar surface field point
to more complex field configurations \citep[e.g.,][]{Donati2010a,
  Donati2011}. In view of this, and noting that the basic idea of flux
trapping, as outlined above, does not depend on the precise field
geometry, \citet{Mohanty2008} generalized the X-wind accretion model
to arbitrary multipole fields.  The fundamental relationship in this
case, for a star of mass $M_{\ast}$, radius $R_{\ast}$ and angular
velocity $\Omega_{\ast}$, is
$$ F_h R_{\ast}^2 \bar{B}_h = \bar{\beta}f^{1/2}(G M_{\ast} \dot{M}_D / \Omega_{\ast})^{1/2} \eqno(1).$$
Here $F_h$ is the fraction of the surface area $2\pi R_{\ast}^2$ of
one hemisphere of the stellar surface (either above or below the
equatorial plane) covered by accretion hot spots with mean field
strength $\bar{B}_h$; $\bar{\beta}$ is a dimensionless, inverse
mass-loading parameter, that measures the ratio of magnetic field to
mass flux; and $f$ is the fraction of the total disk accretion rate
$\dot{M}_D$ that flows into the wind (so 1-$f$ is the fraction that
accretes onto the star). Equation (1) encapsulates the concept of flux
trapping: it relates the amount of observed flux in hot spots on the
 left-hand side (which equals the amount of trapped flux in the X-region that
is loaded with infalling gas) to independently observable quantities
on the right-hand side, without any assumptions about the specific geometry of
the stellar field that ultimately drives the funnel flow.

There is now some significant evidence for the generalized X-wind
model. First, surveys of classical T Tauri stars in Taurus and NGC 2264 strongly support
the correlation $F_h R_{\ast}^2 \propto (G M_{\ast} \dot{M}_D /
\Omega_{\ast})^{1/2}$, predicted by equation (1) if $\bar{B}_h$ can be
assumed constant, an assumption admittedly open to debate
\citep{Johns-Krull2002, Cauley2012}. Second, in two stars with
directly measured $\bar{B}_h$ from spectropolarimetric data, as well
as estimates of $F_h$, $\dot{M}$ and stellar parameters \citep[V2129
Oph and BP Tau;][ respectively]{Donati2007, Donati2008},
\citet{Mohanty2008} find that equation (1) (with some simplistic
assumptions about $\beta$ and $f$; see also below) produces excellent
agreement with the observed $\bar{B}_h F_h$; they also find that
numerical models incorporating the mix of multipole components
observed on these stars give $F_h$, $\bar{B}_h$ and hot spot latitudes
consistent with the data. More generally, there is substantial
evidence that disks are involved in regulating the angular momentum of
the central star, but evidence for disk-locking (disk truncation close
to the corotation radius $R_X$), as specifically proposed by X-wind
theory, is as yet inconclusive; more detailed studies of statistical
significant samples are required (see Section~2.1).

Support for the X-wind model from numerical simulations has so far
been mixed. Using dipole stellar fields, \citet{Romanova2007} have
obtained winds and funnel flows consistent with the theory over
extended durations, but many others \citep[e.g.,][]{Long2007,
  Long2008} have failed. It is noteworthy in this regard that, in the
presence of finite resistivity $\eta$, the flux trapping that is key
to the X-wind model requires that field diffusion out of the X-region
be offset by fluid advection of field into it. This in turn demands
that $\eta/\nu \ll 1$, where $\nu$ is the disk viscosity
\citep{Shu2007}. \citet{Romanova2007} explicitly show that
this condition is critical for attaining an X-type magnetic
configuration, while other simulations typically have $\nu \sim \eta$
instead.

Finally, as an ideal-MHD, steady-state, axisymmetric semi-analytic
model with a stellar field aligned with the rotation axis, X-wind
theory clearly has limitations. For instance, with non-zero
resistivity, reconnection events can lead to episodic outbursts;
similarly, changes in the stellar field or disk accretion rate, and
tilted and/or non-axisymmetric fields can all lead to temporally
varying phenomena, as indeed observed in classical T Tauri stars
\citep[e.g.,][]{Alencar2012}. X-wind theory can at best represent only
the time-averaged behaviour of an intrinsically time-variable system;
numerical simulations are essential for quantitatively exploring these
issues in detail \citep[e.g.,][]{Romanova2011, Long2011}. Similarly,
in the idealized model, the X-region shrinks to a mathematical point,
and how the infinitesimal $\nu$ and $\eta$ then load field lines is
not answerable within the theory, leading to somewhat ad hoc estimates
for $f$ and $\beta$ \citep[see][]{Mohanty2008, Cai2008}. An associated
question is what the appropriate ratio $\eta/\nu$ is for realistic
disks. Global numerical simulations of the magneto-rotational
instability (MRI), with non-zero magnetic flux, are vital to shed
light on these issues \citep[see][]{Shu2007}.

\bigskip
\noindent
\textbf{ 3.2 Stellar winds}
\bigskip

Most stars spend the majority of their lives in isolation, in a sense
that after $\sim$10~Myr of age, they are no longer accreting material
and most do not posses companions that are within reach of their
magnetospheres nor close enough for significant tidal interactions.
For isolated stars, the only available way of losing substantial
angular momentum is by losing mass.  It has been known for a long time
that low-mass stars (those with substantial convective envelopes) are
magnetically active and spin-down via stellar winds
\citep[e.g.,][]{parker58, Kraft:1967p4433, Skumanich:1972p4350,
  Soderblom:1983p4391}.  The coupling of the magnetic field with the
wind can make the angular momentum loss very efficient, in a sense
that the fractional loss of angular momentum can be a few orders of
magnitude larger that the fractional amount of mass lost
\citep[e.g.,][]{Schatzman:1962p4464, Weber:1967p3752,
  Mestel:1968p4497, Reiners:2009p5147}.

In order to calculate the amount of angular momentum loss due to
magnetized stellar winds, the theory generally assumes the conditions
of ideal MHD and a steady-state flow.  These assumptions are acceptable
for characterizing the average, global wind properties, as needed to
understand the long-timescale evolution of stellar rotation.  In this
case, the torque on the star can generally be expressed as
\citep[e.g.,][]{Matt2012b}
\begin{eqnarray}
\label{eq_tausw}
T_{\rm w} = K \; (2 G M_*)^{-m} \; R_*^{5m+2} \; \dot
M_{\rm w}^{1-2m} \; B_*^{4m} \; \Omega_*,
\end{eqnarray}
where $M_*$ and $R_*$ are the stellar mass and radius, $\dot M_{\rm w}$
the mass loss rate, $B_*$ the stellar magnetic field, and $\Omega_*$
the stellar angular velocity. $K$ and $m$ are dimensionless numbers
that depend upon the interaction between the accelerating flow and
rotating magnetic field.

Until recently, most models for computing the angular momentum
evolution of stars (discussed below, see Section 4) have used the
stellar wind torque formulation of \citet[][based on
\citealp{Mestel:1984p2919}]{Kawaler1988}, which is equivalent to
equation (\ref{eq_tausw}) with a value of $m = 0.5$ and fitting the
constant $K$ in order to match the present day solar rotation rate.
This formulation is convenient because, for $m = 0.5$, the stellar
wind torque is independent of the mass loss rate (see eq.\
(\ref{eq_tausw})).  For given stellar parameters, one only needs to
specify how the surface magnetic field strength depends upon rotation
rate, discussed further below.

However, Kawaler's formulation relies upon a 1D approach and adopts
simple power-law relationships for how the magnetic field strength and
the wind velocity vary with distance from the star.
\citet{mattpudritz08II} pointed out that these assumptions are not
generally valid in multi-dimensional winds and that numerical
simulations are needed to accurately and self-consistently determine
the values of $K$ and $m$.  Using 2D (axisymmetric) numerical MHD
simulations, \citet{mattpudritz08II} and \citet{Matt2012b} computed
steady-state solutions for coronal (thermally driven) winds in the
case of stars with a dipolar magnetic field aligned with the rotation
axis.  Using parameter studies to determine how the torque varies,
\citet{Matt2012b} found
\begin{eqnarray}
\label{eq_m12}
m \approx 0.2177  \;\;\;\;\;\;\;\; 
K \approx 6.20 \; [1 + (f / 0.0716)^2]^{-m},
\end{eqnarray}
where $f$ is the stellar rotation rate expressed as a fraction of
breakup (Keplerian rate at the stellar surface),
$f^2~\equiv~\Omega_*^2R_*^3(G M_*)^{-1}$.  Note that the dimensionless
factor $K$ now contains a dependence on the stellar spin rate (in
addition to that appearing in eq.\ (\ref{eq_tausw})), and it is nearly
constant when the star rotates slower than a few percent of breakup
speed.

The formulation of equations (\ref{eq_tausw}) and (\ref{eq_m12}) is
derived from simulations with fixed assumptions about the wind driving
(e.g., coronal temperature) and a particular (dipolar) field geometry
on the stellar surface.  Thus, further work is needed to determine how
the torque depends upon the wind driving
properties\footnote{\citet{uddoula3ea09} reported $m \approx 0.25$,
  derived from their simulations of massive star winds, driven by
  radiation, rather than thermal pressure.  This suggests that the
  value of $m$ does not strongly vary with the wind driving
  properties.} and varies for more complex magnetic geometries.
However, this is the most-dynamically self-consistent stellar wind
torque formulation to date.

After adopting equation (\ref{eq_m12}), it is clear that the torque in
equation (\ref{eq_tausw}) depends upon both the surface magnetic field
strength and the mass loss rate.  To specify the magnetic field
strength, most spin evolution models adopt the relationship suggested
by \citet{Kawaler1988}, $B_* \propto \Omega_*^a R_*^{-2}$.  The
value of $a$ is usually taken to be unity for slow rotators, but above
some critical rotation rate of approximately 10~$\Omega_\odot$ for
solar-type stars, the magnetic field is
taken to be independent of rotation rate by setting $a=0$. The 
so-called ``dynamo saturation'' above some critical velocity is
observationally supported both by direct magnetic field measurements
\citep[e.g.,][]{Reiners:2009p538} and activity proxies \citep[e.g.,][]{Wright:2011p48}.

Recently, \citet{Reiners2012} pointed out that both observations
of magnetic fields \citep[e.g.,][]{Saar:1996p5168, Reiners:2009p538}
and dynamo theory \citep[e.g.,][]{Durney1972, Chabrier2006} were more
consistent with a dynamo relationship following $B_* \propto
\Omega_*^a$---that is, the average magnetic field {\it strength} goes
as some power of $\Omega$, instead of the magnetic {\it flux} (as in
the Kawaler formulation). This has important implications for the
dependence of the torque on the mass (and radius) of the star.
Furthermore, \citet{Reiners2012} derived a new torque
formulations, based on similar assumptions to Kawaler's, arriving at
the equivalent of equation (\ref{eq_tausw}), with $m=2/3$.  Although
this value of $m$ is inconsistent with the MHD simulation results
discussed above, \citet{Reiners2012} demonstrated that, in order
to simultaneously explain the observed spin evolution of both solar
mass and very low mass stars, the stellar wind torque must depend much
more strongly on the stellar mass (and radius) than the Kawaler
formulation.

To calculate the torque, we also need to know how the mass loss rate
depends on stellar properties.  Due to the relatively low loss rates
of non-accreting sun-like and low mass stars, observational detections
and measurements are difficult.  So far, most of what we know is based
on approximately a dozen measurements by \citet{Wood:2002p1260,
  Wood:2005p1261}, which suggest that the mass loss rates vary nearly
linearly with the X-ray luminosity, up to a threshold X-ray flux,
above which the mass loss rates saturate.  More recently,
\citet{Cranmer:2011p3830} have developed a theoretical framework for
predicting the mass loss rates of low mass stars, based upon the
propagation and dissipation of Alfv\'en waves \citep[and
see][]{Suzuki:2012p5032}.  The models of \citet{Cranmer:2011p3830} and
\citet{Suzuki:2012p5032}
compute the mass loss rate as a function of stellar parameters in
a way that is self-consistent with the scaling of magnetic field
strength with stellar rotation rate (including both saturated and
non-saturated regimes).  These models can now be used, in conjunction
with equations (\ref{eq_tausw}) and (\ref{eq_m12}) to compute the
stellar wind torque during most of the lifetime of an isolated,
low-mass star.

As another means of probing the mass loss rates,
\citet{Aarnio:2012p5028} used an observed correlation between coronal
mass ejections and X-ray flares, together with the observed flare rate
distributions derived from T Tauri stars in Orion
\citep{Albacete2007}, to infer mass loss rates due to
coronal mass ejections alone.  \citet{Drake:2013p5063} presented a
similar analysis for main sequence stars.  Furthermore,
\citet{Aarnio:2012p5028} explored how coronal mass ejections may
influence the angular momentum evolution of pre-main-sequence stars.
They concluded that they are not likely to be important during the
accretion phase or during early contraction, but they could
potentially be important after $\sim 10$ Myr.  Although there are a
number of uncertainties associated with estimating CME mass loss rates
from observed X-ray properties, this an interesting area for further
study.

\bigskip
\noindent
\textbf{3.3 Internal processes}
\bigskip

As angular momentum is removed from the stellar surface by external
processes, such as star-disk interaction and stellar winds, the
evolution of the surface rotation rate depends in part on how
angular momentum is transported in the stellar interior. Two limiting
cases are i) solid-body rotation, where it is assumed that angular
momentum loss at the stellar surface is instantaneously redistributed
throughout the whole stellar interior, i.e., the star has uniform
rotation from the center to the surface, and ii) complete
core-envelope decoupling, where only the outer convective zone is
spun down while the inner radiative core accelerates as it develops
during the PMS, thus yielding a large velocity gradient at the
core-envelope interface. Presumably, the actual rotational profile of
solar-type and low mass stars lies between these two extremes. Until
recently, only the internal rotation profile of the Sun was known
\citep{Schou1998}. Thanks to Kepler data, internal rotation has now
been measured for a number of giant stars evolving off the main
sequence from the seismic analysis of mixed gravity and pressure
  modes \citep[e.g.,][]{Deheuvels2012}. The internal rotation profile
of PMS stars and of field main sequence stars is, however, still largely
unconstrainted by the observations.

A number of physical processes act to redistribute angular momentum
throughout the stellar interior. These include various classes of
hydrodynamical instabilites \citep[e.g.,][]{Decressin2009,
  Pinsonneault2010, Lagarde2012, Eggenberger2012}, magnetic fields
\citep[e.g.,][]{Denissenkov2007, Spada2010, Strugarek2011}, and
gravity waves \citep{Charbonnel2013, Mathis2013}, all of which may be
at work during PMS evolution. At the time of writing this review, only
few of the current models describing the angular momentum evolution of
young stars include these processes from first principles
\citep[e.g.,][]{Denissenkov2010a, Turck-Chieze2010, Charbonnel2013,
  Marques2013, Mathis2013} and highlight the need for additional
physics to account for the observations. Pending fully-consistent
physical models, most current modeling efforts adopt empirical
prescriptions for core-envelope angular momentum exchange as discussed
below.

 \bigskip

 \centerline{\textbf{ 4. ANGULAR MOMENTUM EVOLUTION MODELS}}
\bigskip

In an attempt to account for the obsevational results described in
Section~2, most recent models of angular momentum evolution rest on
the 3 main physical processes described in Section~3, namely:
star-disk interaction, wind braking and angular momentum
redistribution in the stellar interior. Each of these
processes is included in the models in a variety of ways, as described
below: 

\begin{itemize}  

\item {\it Star-disk interaction:} only few recent models attempt to
  provide a physical description of the angular momentum exchange
  taking place between the star and its accretion disk. For instance,
  \citet{Matt2012a} computed the evolution of the torque exerted by
  accretion-powered stellar winds onto the central star during the
  early accreting PMS phase. Another example, is the work of {\it
    Gallet and Zanni (in prep.)}, who combined the action of
  accretion-driven winds and magnetospheric ejections to account for
  the nearly constant angular velocity of young PMS stars in spite of
  accretion and contraction. Both models require dipolar magnetic
  field components of about 1-2~kG, i.e., on the high side of the
  observed range of magnetic field strength in young stars
  \citep{Donati2009, Donati2013, Gregory2012}.  Most
  other models, however, merely {\it assume} contant angular velocity
  for the central star as long as it accretes from its disk \citep[as
  originally proposed by][]{Koenigl1991}, with the disk lifetime being
  a free parameter in these models \citep[e.g.,][]{Irwin2007, Irwin2008b,
  Bouvier2008, IrwinBouvier2009, Denissenkov2010b, Reiners2012, Gallet2013}.

\item {\it Wind braking:} up to a few years ago, most models
  used \citeapos{Kawaler1988} semi-empirical prescription, with the
  addition of saturation at high velocities \citep[as originally
  suggested by][]{Stauffer1987}, to estimate the angular momentum loss
  rate due to magnetized winds \citep[see, e.g.,][]{Bouvier1997,
    Krishnamurthi1997, Sills2000}. Recently, more physically-sounded
  braking laws have been proposed. \citet{Reiners2012} revised
  Kawaler's prescription on the basis of a better understanding of
  dynamo-generated magnetic fields, while \citet{Matt2012b} used 2D
  MHD simulations to derive a semi-analytical formulation of the
  external torque exerted on the stellar surface by stellar winds. The
  latter result has been used in the angular momentum evolution models
  developped for solar-type stars by \citet{Gallet2013}, who
  also provide a detailed comparison between the various braking laws.

\item {\it Internal angular momentum transport:} while some models do
  include various types of angular momentum transport processes
  \citep[e.g.,][]{Denissenkov2010a, Turck-Chieze2010, Charbonnel2013,
    Marques2013}, the most popular class of models so far rely on the
  simplifying assumption that the star consists of a radiative core
  and a convective envelope that are both in solid-body rotation but
  at different rates. In these so-called double-zone models, angular
  momentum is exchanged between the core and the envelope at a rate
  set by the core-envelope coupling timescale, a free parameter of
  this class of models \citep[e.g.,][]{Irwin2007, Bouvier2008,
    Irwin2009, Denissenkov2010b, Spada2011}. When dealing with fully
  convective interiors, whether PMS stars on their Hayashi track or
  very-low mass stars, models usually assume solid-body rotation
  throughout the star.

  We illustrate below how these classes of models account for the
  observed spin rate evolution of solar-type stars, low-mass and very
  low-mass stars, and brown dwarfs.

\end{itemize}

\bigskip
\noindent
\textbf{ 4.1 Solar-type stars}
\bigskip

Figure~\ref{gb13} \citep[from][]{Gallet2013} shows the observed and
modeled angular momentum evolution of solar-type stars in the mass
range 0.9-1.1~M$_\odot$, from the start of the PMS at 1~Myr to the age
of the Sun. The rotational distributions of solar-type stars are shown
at various time steps corresponding to the age of the star forming
regions and young open clusters to which they belong
(cf. Fig~\ref{pdist}). Three models are shown, which start with
initial periods of 10, 7, and 1.4~days, corresponding to slow, median,
and fast rotators, respectively. The models assume constant angular
velocity during the star-disk interaction phase in the early PMS,
 and implement the \citet{Matt2012b} wind braking prescription, as well as
core-envelope decoupling. The free parameters of the models are the
initial periods, chosen to fit the rotational distributions of the
earliest clusters, the star-disk interaction timescale $\tau_d$ during
which the angular velocity is held constant at its initial value, the
core-envelope coupling timescale $\tau_{ce}$, and the calibration
constant $K_W$ for wind-driven angular momentum losses. The latter is
fixed by the requirement to fit the Sun's angular velocity at the
Sun's age. These parameters are varied until a reasonable agreement
with observations is obtained. In this case, the slow, median, and
fast rotator models aim at reproducing the 25, 50, and 90$^{th}$
percentiles of the observed rotational distributions and their
evolution from the early PMS to the age of the Sun.

This class of models provide a number of insights into the physical
processes at work. The star-disk interaction lasts for a few Myr in
the early PMS, and possibly longer for slow rotators
($\tau_d\simeq$5~Myr) than for fast ones ($\tau_d\simeq$2.5~Myr). As
the disk dissipates, the star begins to spin up as it contracts
towards the ZAMS. The models then suggest much longer core-envelope
coupling timescales for slow rotators ($\tau_{ce}\simeq$30~Myr) than
for fast ones ($\tau_{ce}\simeq$12~Myr). Hence, on their approach to
the ZAMS, only the outer convective envelope of slow rotators is spun
down while their radiative core remains in rapid rotation. They
consequently develop large angular velocity gradients at the
  interface between the radiative core and the convective envelope on
the ZAMS and, indeed, most of their initial angular momentum is then
hidden in their radiative core \citep[cf.][]{Gallet2013}\footnote{
  Note that the effect of hiding some angular momentum in the
  radiative core would "smooth out" the torques shown in
  Fig.~\ref{fig_hypotorque}, where solid-body rotation was
  assumed. Namely, the ZAMS torques will be slightly less and the
  post-ZAMS will be slightly larger to an age of $\sim$1~Gyr, due to
  the effects of differential rotation and core-envelope
  decoupling}. As stars evolve on the early MS, wind braking
eventually leads to the convergence of rotation rates for all models
by an age of $\simeq$1~Gyr. This is due to the strong dependency of
the braking rate onto the angular velocity: faster rotators are braked
more efficiently than slow ones. Also, the early-MS spin evolution of
slow rotators is flatter than that of fast rotators, in part because
the angular momentum hidden in the radiative core at the ZAMS
resurfaces on a timescale of a few 0.1 Gyr on the early MS. These
models illustrate the strikingly different rotational histories
solar-type stars may experience prior to about 1 Gyr, depending mostly
on their initial period and disk lifetime. In turn, the specific
rotational history a young star undergoes may have a long-lasting
impact on its properties, such as lithium content, even long after
rotational convergence is completed \citep[e.g.,][]{Bouvier2008,
  Randich2010}.

\begin{figure*}[t]
\center
\includegraphics[width=0.8\linewidth,angle=0]{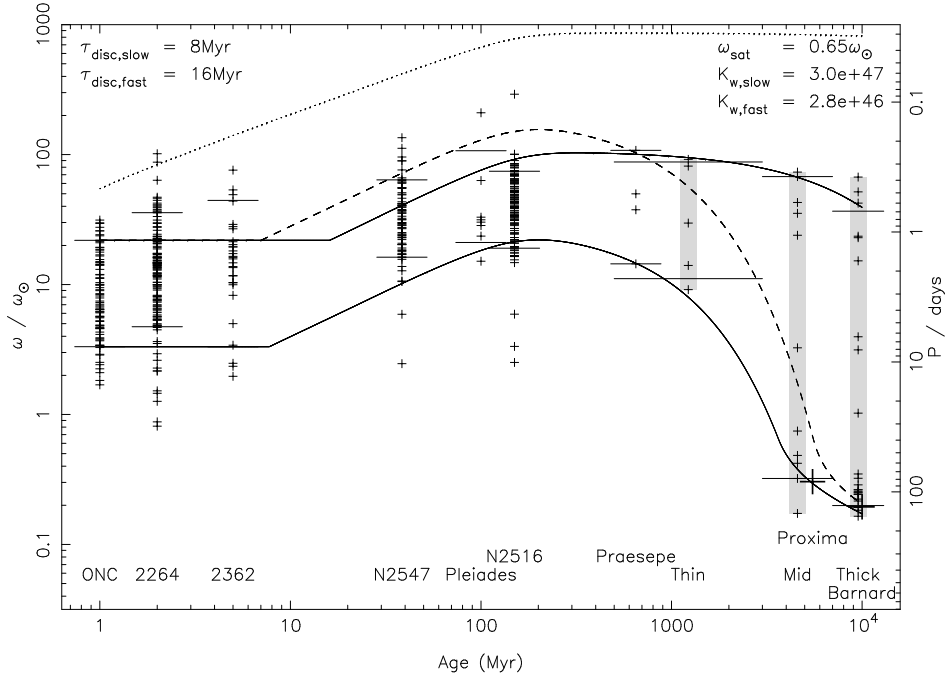}
\caption{\label{irwin11} The rotational angular velocity of very low-mass stars
  (0.1-0.35 M$_\odot$) is plotted as a function of age. The left y-axis is
  labelled with angular velocity scaled to the angular velocity of the
  present Sun while the right y-axis is labelled with rotational
  period in days. On the x-axis the age is given in Myr. {\it Observations:}
  The black crosses shown at various age steps are the rotational
  periods measured for very low-mass stars in star forming regions,
  young open clusters, and in the field over the age range 1 Myr-10
  Gyr. Short horizontal lines show the 10th and 90th percentiles of
  the angular velocity distributions at a given age, used to
  characterize the slow and fast rotators, respectively. {\it Models:} The
  solid curves show rotational evolution models for 0.25 M$_\odot$ stars, fit
  to the percentiles, with the upper curve for the rapid rotators
  (with parameters $\tau_{d,fast}$ and K$_{W,fast}$) and the lower curve for the
  slow rotators (with parameters $\tau_{d,slow}$ and K$_{W,slow}$). Note the factor
  of 10 difference between K$_{W,fast}$ and K$_{W,slow}$. The dashed curve shows
  the result for the rapid rotators if the wind parameter K$_{W,fast}$ is
  assumed to be the same as for the slow rotators rather than allowing
  it to vary. The dotted curve shows the break-up limit. From \citet{Irwin2011}.}
\end{figure*}

These models describe the spin evolution of isolated stars while many
cool stars belong to multiple stellar systems
\citep[cf.][]{Duchene2013}. For short period binaries
($P_{orb}\leq$12~days), tidal interaction enforces synchronization
between the orbital and rotational periods \citep{Zahn1977} and the
spin evolution of the components of such systems will clearly differ
from that of single stars \citep{Zahn1989}.  However, the fraction of
such tight, synchronized systems among solar-type stars is low, of
order of 3\% \citep{Raghavan2010}, so that tidal effects are unlikely
to play a major role in the angular momentum evolution of most cool
stars.  Another potentially important factor is the occurrence of
planetary systems \citep[e.g.,][]{Mayor2011, Bonfils2013}. The
frequency of hot Jupiters, i.e., massive planets close enough to their
host star to have a significant tidal or magnetospheric influence
\citep[cf.][]{Dobbs-Dixon2004, Lanza2010, Cohen2010}, is quite low,
amounting to a mere 1\% around FGK stars
\citep[e.g.,][]{Wright2012}. However, there is mounting evidence that
the formation of planetary systems is quite a dynamic process, with
gravitational interactions taking place between forming and/or
migrating planets \citep[][see also the chapters by {\it Davies et
  al.}  and {\it Baruteau et al.}]{Albrecht2012}, which may lead to
planet scattering and even planet engulfment by the host star. The
impact of such catastrophic events onto the angular momentum evolution
of planet-bearing stars has been investigated by \citet{Bolmont2012}
who showed it could significantly modify the instantaneous spin rate of
planet host stars both during the PMS and on the main sequence.

\bigskip
\bigskip
\bigskip
\noindent
\textbf{ 4.2 Very low-mass stars}
\bigskip

\begin{figure*}[t]
\center\includegraphics[width=0.8\linewidth]{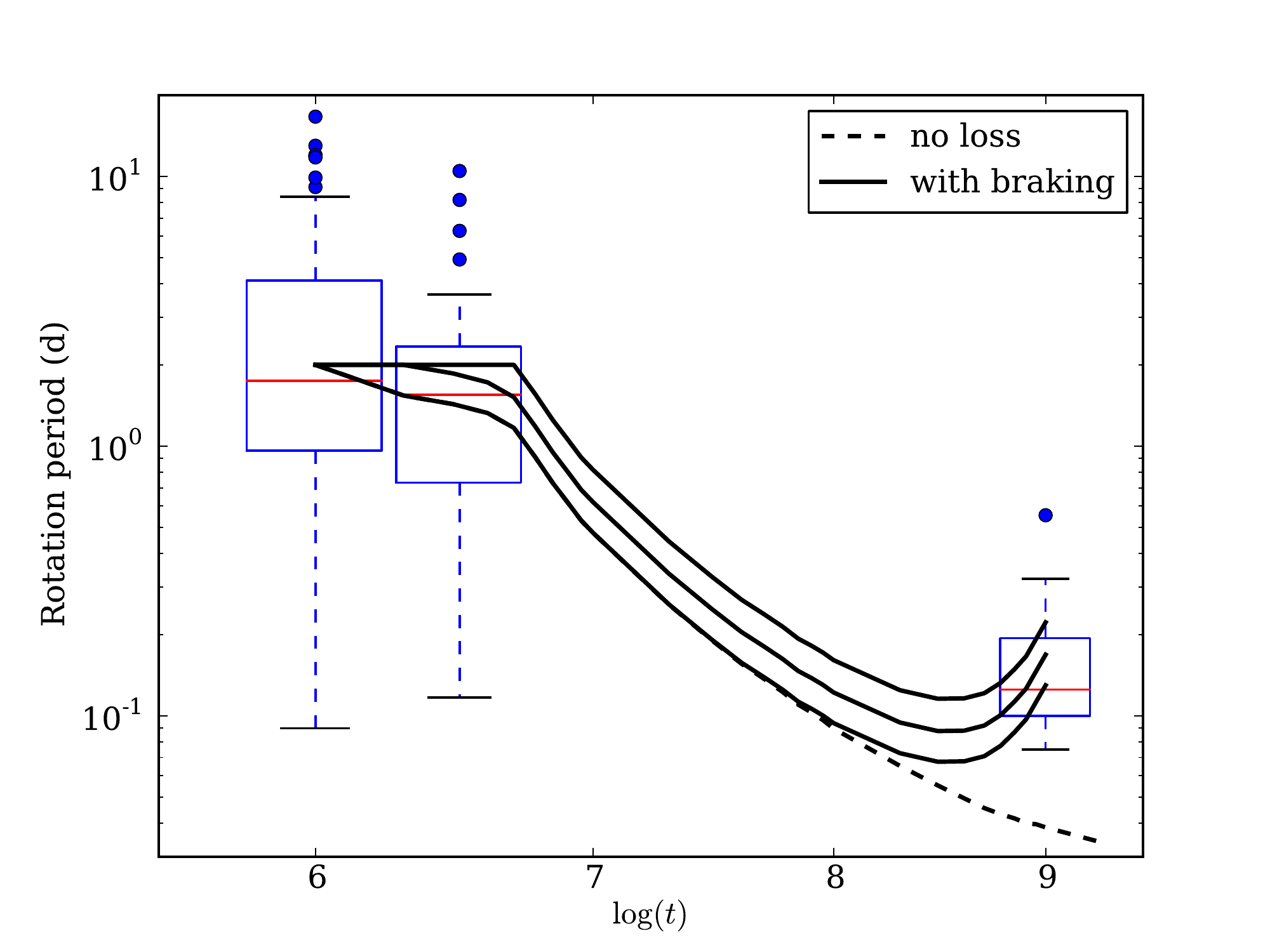}
\caption{\label{bdjevol}  Boxplot showing rotation periods for brown
  dwarfs as a function of age. The plot contains periods for
  $\sim$1~Myr \citep[ONC][]{Rodriguez2009}, for $\sim$3~Myr
  \citep{Cody2010, Scholz2004a, Scholz2005, Bailer-Jones2001,
    Zapatero2003, Caballero2004}, and for the field population, for
  convenience plotted at an age of 1\,Gyr although individual ages may
  vary \citep{Bailer-Jones2001, Clarke2002, Koen2006, Lane2007,
    Artigau2009, Radigan2012, Girardin2013, 
    Gillon2013, Heinze2013}. {\it Red line:} median;
    {\it box:} lower and upper quartile; {\it 'whiskers':} range of
    datapoints within 1.5$\times$ (75\% - 25\%) range. Outliers
    outside that range are plotted as individual datapoints. Note that
    a few more brown dwarf periods have been measured at other ages,
    not shown here due to the small sample sizes. The dashed line
    illustrates evolution models without angular momentum loss. The 3
    solid lines correspond to models including saturated angular momentum
    losses and disk locking phases lasting
    for 1, 2, and 5\,Myr, respectively, with object radii taken from the
    0.05~M$_\odot$ BT-Settl evolutionary models of 
    \citet{Allard2011}. The best fit to the current observational
    constraints is obtained by assuming an angular momentum loss rate
    for brown dwarfs that is $\sim$10,000 times weaker than that used
    for solar-type
    stars shown in Fig.~\ref{gb13}.}
 \end{figure*}

 Models similar to those described above for solar-type stars have
 been shown to apply to lower mass stars, at least down to the fully
 convective boundary ($\simeq$0.3~M$_\odot$), with the core-envelope
 coupling timescale apparently lengthening as the convective envelope
 thickens \citep[e.g.,][]{Irwin2008a}. In the fully convective regime,
 i.e., below 0.3~M$_\odot$, models ought to be simpler as the
 core-envelope decoupling assumption becomes irrelevant and uniform
 rotation is usually assumed throughout the star. Yet, the rotational
 evolution of very low-mass stars actually appears more complex than
 that of their more massive counterparts and still challenges current
 models. Figure~\ref{irwin11} \citep[from][]{Irwin2011} shows that
 disk locking still seems to be required for VLM stars in order to
 account for their slowly evolving rotational period distributions
 during the first few Myr of PMS evolution. Yet, as discussed above
 (see Section 2.2), the evidence for a disk-rotation connection in
 young VLM stars is, at best, controversial. Equally problematic, the
 rotational period distribution of field M-dwarfs appears to be
 bimodal, with pronounced peaks at fast (0.2-10~d) and slow (30-150~d)
 rotation \citep{Irwin2011}. Most of the slow rotators appear to be
 thick disk members, i.e., they are on average older than the fast
 ones that are kinematically associated to the thin disk, and the
 apparent bimodality coud thus simply result from a longer spin down
 timescale for VLM stars, of order of a few Gyr, as advocated by
 \citet{Reiners2012} and \citet{McQuillan2013}.

 However, as shown in Figure~\ref{irwin11}, this bimodality may not be
 easily explained for field stars at an age of several Gyr.  Indeed,
 contrary to solar-type stars whose rotational {\it scatter} decreases
 from the ZAMS to the late-MS (cf. Fig.~\ref{gb13}), the distribution
 of spin rates of VLM stars widens from the ZAMS to later ages. The
 large dispersion of rotation rates observed at late ages for VLM
 stars thus requires drastically different model
 assumptions. Specifically, for a given model mass (0.25~M$_\odot$ in
 Fig.~\ref{irwin11}), the calibration of the wind-driven angular
 momentum loss rate has to differ by one order of magnitude between
 slow and fast rotators \citep{Irwin2011}. Why does a fraction of VLM
 stars remain fast rotators over nearly 10~Gyr while another fraction
 is slowed down on a timescale of only a few Gyr is currently
 unclear. A promising direction to better understand the rotational
 evolution of VLM stars is the recently reported evidence for a
 bimodality in their magnetic properties. Based on spectropolarimetric
 measurements of the magnetic topology of late M dwarfs obtained by
 \citet{Morin2010}, \citet{Gastine2013} have suggested that a bistable
 dynamo operates in fully convective stars, which results in two
 contrasting magnetic topologies: either strong axisymmetric dipolar
 fields or weak multipolar fields. Whether the different magnetic
 topologies encountered among M dwarfs is at the origin of their
 rotational dispersion at late ages remains to be assessed.

\bigskip
\bigskip

\bigskip
\noindent
\textbf{ 4.3 Brown dwarfs}
\bigskip

Figure~\ref{bdjevol} illustrates the current rotational data and
models for brown dwarfs (BDs) from 1 Myr to the field substellar
population. As discussed in Section 2.2 above, substellar objects are
characterized by fast rotation from their young age throughout their
whole evolution, with a median period of about 2~d at 1~Myr and 3-4~h
at 1~Gyr. Somewhat controversial evidence for disk locking has been
reported among young BDs (see Sect. 2.2 above), although sensitive
mid-IR surveys are still needed for large samples in order to better
characterize the disk frequency. Figure~\ref{bdjevol} shows models
computed with and without angular momentum losses from (sub)stellar
winds. The evolution of substellar rotational distributions in the
first few Myr is consistent with either no or moderate disk locking,
as previously advocated by \citet{Lamm2005}. At an age of a few Myr,
the observed rotation rates suggest substellar objects experience
little angular momentum loss on this timescale. By an age of a few
Gyr, however, some angular momentum loss has occurred. The best fit to
the observational constraints is obtained with models featuring an
angular momentum loss rate for BDs that is about 10,000 times weaker
than that assumed for solar-type stars (cf. Fig.~\ref{gb13}). Whether
the unefficient rotational braking of brown dwarf results from a
peculiar magnetic topology, their predominantly neutral atmospheres,
or some other cause is currently unclear.

\bigskip
\noindent
{\bf 4.4 Summary} 
\bigskip

Current models of the spin evolution of low-mass stars appear to
converge towards the following consensus:

\begin{itemize}

\item At all masses, the initial distribution of angular momentum
  exhibit a large dispersion that must reflect some process operating
  during the core collapse and/or the embedded protostellar
  stage. Current models do not solve for this initial rotational
  scatter but adopt it as initial conditions.

\item Some disk related process is required at least for
  solar-type and low mass stars during the first few Myr in order to
  account for their hardly evolving rotational period distributions
  during the early PMS. Whether this process is still instrumental in
  the VLM and substellar regimes remains to be assessed. The disk
  lifetimes required by angular momentum evolution models are
  consistent with those empirically derived from the evolution of IR
  excess in young stars \citep[e.g.,][]{Bell2013}.

\item Rotational braking due to magnetized winds is strongly mass
  dependent, being much less efficient at very low masses. At a given
  mass, angular momentum loss must also scale with the spin rate in
  order to account for the rotational convergence of solar-type and
  low-mass stars on a timescale of a few 0.1 Gyr. The spin down
    timescale from the ZAMS increases towards lower mass stars
    (from $\sim$0.1 Gyr at 1 M$_\odot$ to $\sim$1 Gyr at 0.3 M$_\odot$, and
    $\ge$10 Gyr at $\le$0.1 M$_\odot$), but once completed, the
    rotational convergence usually occurs at a lower spin rate for
    lower mass stars \citep{McQuillan2013}.

\item Some form of core-envelope decoupling must be introduced in the
  models in order to simultaneously account for the specific spin
  evolution of initially slow and fast rotators. The
  empirically-derived core-envelope coupling timescale is found to be
  longer in slow rotators than in fast ones at a given mass, thus
  providing some hints at the underlying physical process responsible
  for angular momentum transport in stellar interiors.

\end{itemize}

\bigskip

\centerline{\textbf{ 5. CONCLUSION}}
\bigskip

In the last few years, we have reached a stage where the rotational
evolution of cool stars and brown dwarfs is relatively well
constrained by the observations. Additional rotational period
measurements for homogeneous and coeval populations are still required
to fill a few age and mass gaps, e.g., old ($\geq$1~Gyr) field dwarfs
and ZAMS cluster (0.1-0.5~Gyr) brown dwarfs, so as to provide a
complete picture of the spin evolution of stars and substellar
objects. The physical description of the mechanisms that dictate the
spin evolution of cool stars has also tremendously progressed over the
last years, with the exploration of new processes and the refinement
of prior ones. Yet, the slow rotation rates of young stars still
remain very much of a challenge to these models. A better
characterization of the critical quantities involved in the star-disk
interaction and in stellar winds, such as the stellar magnetic field
intensity and topology, the mass accretion rate onto the star, and the
amount of mass loss a star experiences during its lifetime, is sorely
needed in order to progress on these issues.  In spite of these
limitations, the semi-empirical angular momentum evolution models
developped to date appear to grasp some of the major trends of the
observed spin evolution of cool stars and brown dwarfs. Undoubtly, the
main area of progress to be expected in the next few years lies in the
improved physical modeling of these processes.

\bigskip \bigskip {\it Acknowledgments} The authors would like to
dedicate this contribution to the memory of Jean Heyvaerts who passed
away the week before PPVI. JB acknowledges the grant ANR 2011 Blanc
SIMI5-6 020 01 ”Toupies: Towards understanding the spin evolution of
stars” (http://ipag.osug.fr\-/Anr\_Toupies/ ).  KGS and SPM
acknowledge the USA National Science Foundation (NSF) grant
AST-0808072. KGS also acknowledges NSF grant AST-0849736. SM
acknowledges the support of the UK STFC grant ST/H00307X/1, and many
invaluable discussions with F. Shu and A. Reiners.  We thank Florian
Gallet and Jonathan Irwin for providing Fig.6 and Fig.7 of this review
chapter, respectively. We are indebted to the many authors who have
provided us with the data used to build Fig.1 of this review
chapter.\bigskip \bigskip

\bibliography{sean,jerome,aleks,subu,claudio,keivan}
\bibliographystyle{ppvi_lim3}

\end{document}